\documentclass[a4paper,11pt]{article}
\pdfoutput=1 

\usepackage{jcappub} 

\usepackage[T1]{fontenc}

\title{\boldmath Sensitivity of the Global 21-cm Signal to Dark Matter-Baryon Scattering}

\author[a,1]{Aryan Rahimieh,}
\author[a,2]{Priyank Parashari,}
\author[a]{Rui An,}
\author[a]{Trey Driskell,}
\author[b,c]{Jordan Mirocha,}
\author[a]{and Vera Gluscevic}

\affiliation[a]{Department of Physics and Astronomy, University of Southern California, Los Angeles, CA, 90089, USA}
\affiliation[b]{Jet Propulsion Laboratory, 4800 Oak Grove Drive, Pasadena, CA 91109, USA}
\affiliation[c]{California Institute of Technology, 1200 E. California Boulevard, Pasadena, CA 91125, USA}

\emailAdd{rahimieh@usc.edu}
\emailAdd{ppriyank@usc.edu}
\emailAdd{anrui@usc.edu}
\emailAdd{gdriskel@usc.edu}
\emailAdd{jordan.mirocha@jpl.nasa.gov}
\emailAdd{gluscevi@usc.edu}

\abstract{With current and upcoming experiments on the horizon, the global 21-cm signal can open up new avenues for probing dark matter (DM) physics at redshifts that are otherwise inaccessible to other observables. This work investigates the effects of elastic scattering between DM and baryons on the global 21-cm signal in two distinct interacting DM (IDM) models: Coulomb-like and velocity-independent interactions. Our analysis incorporates key astrophysical parameters essential for accurately modeling the global signal, including star formation efficiency, escape fraction of ionizing photons, normalization of the X-ray luminosity, the number of Lyman-Werner photons emitted per stellar baryon, the minimum virial temperature of star-forming halos, as well as the IDM particle mass and cross section. We perform a Fisher analysis to forecast the sensitivity of four global 21-cm signal experimental scenarios as probes of DM-baryon scattering. We find that global signal experiments, even at the sensitivity of the current facilities such as EDGES and SARAS3, could improve existing cosmological and astrophysical constraints on DM-baryon scattering. Our results also highlight the degeneracies among the DM-baryon interaction cross section and astrophysical quantities. In particular, degeneracies between the IDM cross section and two astrophysical parameters, the minimum virial temperature, and Lyman-Werner photon production, can significantly impact the DM interaction inference. Conversely, the velocity-independent cross section is found to be insensitive to uncertainties in the X-ray luminosity. These findings underscore the necessity of accurately characterizing the uncertainties in astrophysical parameters to leverage the full potential of the 21-cm global signal experiments in probing IDM physics.}

\begin{document}
\maketitle
\flushbottom

\section{Introduction}\label{sec:intro_global}
Many observational probes testify to the existence of Dark Matter (DM), which comprises around $\sim$ 85\% of the matter content in the universe at present~\cite{zwicky2009republication, rubin1970rotation, refregier2003weak, clowe2006direct, aghanim2020planck}. Despite numerous lines of evidence for the existence of DM, its fundamental nature still remains elusive. In standard cosmology, DM is typically modeled as a cold, non-relativistic, collisionless particle, often referred to as cold dark matter (CDM). However, CDM may not be able to account for the full range of observational anomalies reported on various scales \cite{abdalla2022cosmology, anchordoqui2021dissecting, di2021realm, schoneberg2022h0, perivolaropoulos2022challenges, tulin2018dark, zentner2022critical}, which highlights the importance of exploring alternative models beyond CDM. We focus on interacting DM (IDM) \cite{sigurdson2004dark, boehm2005constraints, xu2018probing, dvorkin2014constraining, ooba2019cosmological, slatyer2018early, boddy2018critical, lin2023dark, boddy2018first, gluscevic2018constraints, maamari2021bounds, rogers2022limits, nadler2019constraints, nadler2021constraints, ali2015constraints, nguyen2021observational, slatyer2009cmb, xu2021constraints, becker2021cosmological, boddy2022investigation, li2023atacama, chen2002cosmic, munoz2015heating, mosbech2023probing, fialkov2018constraining, barkana2018possible, short2022dark, driskell2022structure, he2023s8, he2025bounds, mcdermott2011turning, kovetz2018tighter, munoz2018insights, berlin2018severely, barkana2018signs, liu2018implications, munoz201821, liu2019reviving, barkana2023anticipating, escudero2018fresh, creque2019direct}, which features elastic scattering with the Standard Model particles. Significant efforts have been made to search for signatures of IDM in data from a variety of probes~\cite{gluscevic2019cosmological, buen2022cosmological}, such as direct detection experiments of DM~\cite{slatyer2009cmb, aprile2016low, agnes2018constraints, agnese2018first, barak2020sensei}, cosmic rays~\cite{bringmann2019novel, cappiello2019strong}, cosmic microwave background (CMB) observations~\cite{ali2015constraints, gluscevic2018constraints, xu2018probing, nguyen2021observational, slatyer2009cmb}, the Lyman-$\alpha$ forest flux~\cite{dvorkin2014constraining, xu2018probing, ooba2019cosmological}, and Milky Way satellite abundance~\cite{maamari2021bounds, nguyen2021observational}. Complimentary to these experiments, observations of the 21-cm signal from the high-redshift universe can also provide unique exploration opportunities~\cite{edges_nature, singh2022detection, burns2019dark, adams2023improved, munoz2015heating, mosbech2023probing}. IDM models, especially those with Coulomb-like interactions, have garnered significant attention due to their potential to alter the global 21-cm signal from neutral hydrogen at cosmic dawn. Motivated by the distinct effect of IDM on the global 21-cm signal, we focus this study on quantifying the potential of 21-cm cosmology as a probe of IDM. 

A global 21-cm signal originating from neutral hydrogen at the cosmic dawn and the epoch of reionization is a powerful cosmological probe, pursued by a number of experiments, including The Experiment to Detect the Global EoR Signature (EDGES)~\cite{bowman2008toward, bowman2010lower}, which reported~\cite{edges_nature} an absorption feature centered at 78 MHz in the global 21-cm signal. However, the Shaped Antenna measurement of the background RAdio Spectrum (SARAS)~\cite{singh2017first} collaboration subsequently reported a non-detection of the signal at 95.3\% confidence~\cite{singh2022detection}. The report of the EDGES anomaly gained attention, and several explanations for the unusual observed absorption feature in this signal were proposed. These include challenges with foreground removal, systematic errors in the measurements~\cite{hills2018concerns, singh2019redshifted, sims2019joint, sims2020testing, bradley2019ground, tauscher2020global}, as well as new physics beyond the standard cosmological framework, such as IDM scattering with baryons~\cite{barkana2018possible, fialkov2018constraining, kovetz2018tighter, driskell2022structure}, excess radio background from DM decay~\cite{mondal2024constraining}, and modified dispersion relation~\cite{das2022modified}. A variety of different global 21-cm signal experiment designs have been explored over the years, including EDGES, SARAS, the Probing Radio Intensity at High-Z from Marion (PRIZM)~\cite{philip2019probing}, The All-Sky SignAl Short-Spacing INterferometer (ASSASSIN)~\cite{mckinley2020all}, Radio Experiment for the Analysis of Cosmic Hydrogen (REACH)~\cite{de2019reach}, Sonda Cosmológica de las Islas para la Detección de Hidrógeno Neutro (SCI-HI)~\cite{voytek2014probing}, Broadband Instrument for Global HydrOgen ReioNisation Signal (BIGHORNS)~\cite{sokolowski2015bighorns}, the Large-Aperture Experiment to Detect the Dark Ages (LEDA)~\cite{price2018design}, Dark Ages Polarimetry PathfindER (DAPPER)~\cite{burns2021global}, and Discovering the Sky (DSL) \cite{chen2021discovering}. Similar to the global signal, the 21-cm power spectrum is a powerful probe of new physics. In a forthcoming study~\cite{Rahimieh:2025lbf}, we forecast the sensitivity of the Hydrogen Epoch of Reionization Array (HERA)~\cite{deboer2017hydrogen} to IDM models.

This study explores the sensitivity of the global 21-cm experiments to detecting the effects of DM interactions. Following previous literature~\cite{dvorkin2014constraining, nguyen2021observational, becker2021cosmological, li2023atacama, boddy2022investigation, sigurdson2004dark}, we consider IDM models in which the interaction cross section is parameterized as $\sigma(v) = \sigma_\mathrm{0} v^{n}$, where $\sigma_{0}$ is the unknown cross section normalization and $v$ is the relative velocity between DM and baryons. We focus on two models: i) a Coulomb-like interaction ($n=-4$)~\cite{slatyer2018early, boddy2018critical, lin2023dark, driskell2022structure, creque2019direct} with $n=-4$, and ii) a velocity-independent interaction with $n=0$ \cite{boddy2018first, gluscevic2018constraints, maamari2021bounds, rogers2022limits, nadler2019constraints, nadler2021constraints}. To model the global 21-cm signal in presence of DM-baryon scattering, we follow the approach of Ref.~\cite{driskell2022structure}, and include the effects of scattering on structure formation and on the thermal history of baryons, using the semi-analytic merger-tree code $\textsc{Galacticus}$~\cite{benson2012galacticus}, and the 21-cm signal calculator \textsc{Ares}~\cite{mirocha2014}. Using the Fisher matrix formalism, we analyze four distinct experimental scenarios and forecast the sensitivity to recovering the IDM cross section and astrophysical parameters, exploring the degeneracies between them. We find that for the $n=0$ case, a SARAS-like (similar configuration to SARAS3) experiment has a sensitivity comparable to the strongest bounds to date obtained from the Milky Way satellite abundance measurements \cite{maamari2021bounds, nadler2019constraints}. Furthermore, other considered experimental scenarios, which include an EDGES-like experiment (similar configuration to EDGES), Future1 (more integration time compared to EDGES), and Future2 (more integration time and frequency channels compared to EDGES), show stronger sensitivities. For $n=-4$ case, all four scenarios show improvements compared to the strongest bounds obtained from the cosmic microwave anisotropy (CMB) measurements \cite{nguyen2021observational}.

This paper is organized as follows. In Section~\ref{sec:21-cm_model}, we summarize the IDM models and outline our approach for modeling the global 21-cm signal. In Section~\ref{sec:methods_global}, we summarize the Fisher matrix formalism used to calculate the forecasts for various experiment scenarios. In Section~\ref{sec:results_global}, we present the results and discuss their implications. We summarize the main findings of this in Section~\ref{sec:conclusions_global} and briefly discuss the future directions.

Throughout this work, we fix the standard cosmological parameters at their best-fit values inferred from the Planck 2018 measurements~\cite{aghanim2020planck}: $h = 0.6736$, $\Omega_\mathrm{m} = 0.3153$, $\Omega_\mathrm{\Lambda} = 0.6847$, $\Omega_\mathrm{b} = 0.04930$, $T_\mathrm{CMB} = 2.72548$, $n_\mathrm{s} = 0.9649$, $N_\mathrm{eff} = 3.046$, and $\sigma_\mathrm{8} = 0.8111$.

\section{Global 21-cm Signal with IDM}\label{sec:21-cm_model}

We investigate a specific IDM model where DM interacts with baryons through elastic scattering, resulting in the momentum and heat transfer between the corresponding cosmological fluids, and a dissipation of the relative bulk velocity. For non-relativistic scattering, the momentum-transfer cross section can be parameterized as $\sigma(v) = \sigma_\mathrm{0} v^{n}$, where $v$ is the relative velocity between the interacting particles and $\sigma_\mathrm{0}$ is the free fitting parameter of the model, representing the cross section normalization. We consider two scenarios, a Coulomb-like interaction ($n=-4$) and velocity-independent interaction ($n=0$). In both cases, DM particles scatter with neutral hydrogen and helium atoms, following Refs.~\cite{barkana2018possible, liu2019reviving, driskell2022structure}. To model the Global 21-cm signal in the IDM cosmology, we follow the methodology developed in Ref.~\cite{driskell2022structure}. In the following,
we briefly summarize this formalism and refer the reader to Ref.~\cite{driskell2022structure} for a comprehensive discussion.

The interaction between DM and baryon modifies the evolution of perturbations in the DM-baryon fluid, which introduces some modifications to the linear Boltzmann equations for both species. In synchronous gauge, and assuming both fluids are non-relativistic, the perturbed Boltzmann equations take the following form
\begin{align}
\dot{\delta}_\chi &= -\theta_\chi - \tfrac{\dot{h}}{2}, 
\qquad \dot{\delta}_b = -\theta_b - \tfrac{\dot{h}}{2},, \nonumber \\
\dot{\theta}_\chi &= -\frac{\dot{a}}{a}\,\theta_\chi + c_\chi^2 k^2 \delta_\chi + R_\chi(\theta_b - \theta_\chi), \nonumber \\
\dot{\theta}_b &= -\frac{\dot{a}}{a}\,\theta_b + c_b^2 k^2 \delta_b + \frac{\rho_\chi}{\rho_b} R_\chi(\theta_\chi - \theta_b) + R_\gamma(\theta_\gamma - \theta_b),
\label{eq:Boltzmann-mod}
\end{align}
where $\delta_\chi$ and $\delta_b$ denote the DM and baryon density perturbations, $\theta_\chi$ and $\theta_b$ are their velocity divergences, $c_\chi$ and $c_b$ are the respective sound speeds, $h$ is the trace of the scalar metric perturbation, and $R_\gamma$ represents the Compton momentum–transfer rate between baryons and photons. Equation~\eqref{eq:Boltzmann-mod} highlights that the drag terms proportional to $R_\chi$ are the point through which the microscopic interaction parameters $(m_\chi,\sigma_0)$ influence the evolution of cosmological perturbations. For DM scattering off multiple baryonic targets, the coefficient $R_\chi$ is given by
\begin{equation}
R_\chi \;=\; \sum_t N_n \, \frac{a \rho_t \sigma_0}{m_\chi + m_t}
\left(\frac{T_\chi}{m_\chi}+\frac{T_K}{m_t}+\frac{V_{\rm rms}^2}{3}\right)^{\frac{n+1}{2}},
\label{eq:Rchi}
\end{equation}
where $N_n = 2^{(n+5)/2}\Gamma\!\left(3+\frac{n}{2}\right)/(3\sqrt{\pi})$ and the sum is over baryonic targets $t$, with mass $m_t$ and density $\rho_t$, and $T_\chi$, $T_K$, and $V_{\rm rms}$ denote the DM temperature, baryon temperature, and rms bulk relative velocity, respectively. This expression shows explicitly how the DM mass and cross-section affects the efficiency of momentum and heat transfer between the two fluids. Together, Eqs.~\eqref{eq:Boltzmann-mod} and \eqref{eq:Rchi} provide the essential link between the microphysical scattering process and the macroscopic evolution of cosmological perturbations. The DM-baryon interaction also affects the temperature evolution of both the species, thereby influencing the Global 21-cm signal. We use a modified version of the publicly available code \textsc{class}\footnote[1]{\url{https://github.com/kboddy/class_public/tree/dmeff}} \cite{nguyen2021observational, gluscevic2018constraints, boddy2018first}, which self-consistently computes the linear matter power spectrum, evolution of DM and baryon temperatures, and their relative bulk velocity in the presence of DM-baryon interactions. 
As noted in previous literature, in case of a Coulomb-like interaction, the matter power spectrum is affected across a range of scales, with a characteristic scale-dependent suppression, with an onset at $k \approx 10^{-2} \ h \ \rm{Mpc}^{-1}$; on smaller scales, the transfer function shows a decrement of power  (see Figure $3$ in Ref.~\cite{driskell2022structure}), with the suppression inversely proportional to $\sigma_\mathrm{0}$. In the velocity-independent scattering model, the suppression of the matter power spectrum is similar to that of warm DM, and the interactions in the late-time are negligible~\cite{nadler2019constraints}. 

The second step of the formalism involves modeling the nonlinear growth of DM fluctuations and computing the comoving number density per unit mass of DM halos formed, which is referred to as the halo mass function (HMF). In the extended Press–Schechter (ePS) formalism~\cite{press1974formation, bond1991excursion}, the HMF is defined as
\begin{equation}\label{eq:hmf}
	\frac{dn}{dM} = f(\sigma) \frac{\rho_\mathrm{m}}{M} \frac{d \ln(\sigma^{-1})}{d M} \,,
\end{equation}
where $n$ is the comoving number density of halos of mass $M$, $f(\sigma)$ is a fitting function to match HMFs obtained from N-body simulations, $\rho_\mathrm{m}$ is the mean matter density, and the filtered mass variance $\sigma(m)$ is computed as an integral of the linear matter power spectrum times a window function $W(k|M)$:
\begin{eqnarray}
	\sigma^{2}(M) = \frac{1}{2\pi^2} \int^{\infty}_\mathrm{0} 4\pi k^2P(k)W^2(k|M)dk\,.
\end{eqnarray}
We use the sharp-k window function,~\cite{schneider2015structure} as it is sensitive to sharp cutoffs in the power spectrum. This choice ensures that halos are not formed too far below the cut-off scale, which could happen for other window functions like the top-hat filter~\cite{benson2013dark}. Sharp-k filter is defined as:
\[
W(k|m) =
\begin{cases}
	1 & \text{if } k \leq k_\mathrm{s}(M) \\
	0 & \text{if } k > k_\mathrm{s}(M),
\end{cases}
\]
where
\begin{eqnarray}
	k_\mathrm{s}(M)=2.5/R \quad \text{and} \quad R= \left( \frac{3M}{4\pi \bar{\rho}} \right)^{1/3}.
\end{eqnarray}
We also use the Tinker fitting function \cite{tinker2008toward}:
\begin{eqnarray}
	f(\sigma)= A \left[ \left( \frac{\sigma}{b}  \right)^{-a} +1 \right] e^{-c/\sigma^2},
\end{eqnarray}
where $A$ controls the amplitude, $a$ controls the tilt, $b$ sets the mass scale at which the power law becomes significant, and $c$ determines the high-mass cutoff scale above which halo abundances exponentially decrease. This function is widely used, and it has successfully reproduced the HMFs at low z predicted by N-body simulations~\cite{despali2016universality,ondaro2022non}. We use \textsc{galacticus}\footnote[2]{\url{https://github.com/galacticusorg/galacticus}} code, which utilizes the ePS formalism, to calculate the HMF for the given initial conditions generated using \textsc{CLASS} up to $z=500$. We choose $10^5 M_{\odot}$ as the resolution in \textsc{galacticus} in order to capture the details of the smallest scales important for the Global signal. As discussed in Ref.~\cite{driskell2022structure}, the size of the sound horizon is evolving, and due to interactions, the structure formation may be affected, especially at late times. Such interactions can suppress the growth of small-scale structures, leading to a modification in the HMF by reducing the abundance of low-mass halos compared to the standard CDM scenario.

\subsection{21-cm Global Signal}\label{sec:ares}
The redshifted global 21-cm signal is defined as the brightness temperature relative to the CMB temperature \cite{furlanetto2006global}:
\begin{eqnarray} \label{eq:glob21}
    \delta T_\mathrm{b} \simeq &27(1-\bar x_i)\left(\frac{1-Y_p}{0.76}\right) \left(\frac{\Omega_b h^2}{0.023}\right)\left(\frac{0.15}{\Omega_m h^2} \frac{1+z}{10}\right)^{1/2} \left(1-\frac{T_\mathrm{\gamma}}{T_\mathrm{S}}\right) mK, 
\end{eqnarray} 
where $x_i$ is the mean ionized fraction, $Y_\mathrm{p}$ is the Helium mass fraction, $T_S$ is the spin temperature, and $T_\mathrm{\gamma}$ is the CMB temperature. The spin temperature, $T_\mathrm{S}$, in Eq.~\ref{eq:glob21} depends on three competing processes: coupling to the CMB photons, collisional coupling with the gas, and Lyman-$\alpha$ coupling:
\begin{eqnarray}
T_\mathrm{S}^{-1}=\frac{T_\mathrm{\gamma}^{-1}+x_\mathrm{c} T_\mathrm{K}^{-1}+x_\mathrm{\alpha} T_\mathrm{\alpha}^{-1}}{1+x_\mathrm{c}+x_\mathrm{\alpha}},
\end{eqnarray}
where $x_\mathrm{c}$ is the collisional coupling coefficient \cite{Zygelman2005}, $T_\mathrm{K}$ is the kinetic temperature of the gas, and $x_\mathrm{\alpha}$ is the coefficient of Lyman-$\alpha$ coupling with $T_\mathrm{\alpha}$ being the corresponding Lyman-$\alpha$ color temperature. In this work, we use a modified version of \textsc{ares}\footnote[3]{\url{https://github.com/treydriskell/ares/tree/galacticus_hmf}} \cite{mirocha2014} to predict the Global signal utilizing the output obtained from \textsc{galacticus} for a given set of astrophysical and IDM parameters. In the computation of the global 21-cm signal, \textsc{ares} takes into account for the Lyman-$\alpha$ flux, X-ray heating, and ionizing emission effects in the presence of DM-baryon interactions. Key modifications to \textsc{ares} are the gas kinetic temperature and the Lyman-$\alpha$ coupling\footnote[4]{For more details we refer the reader to Ref.~\cite{driskell2022structure}}.

In this work, we use a simple parameterization of astrophysical sources that relates photon production in the UV and X-ray bands to the rate at which baryons collapse into DM halos. The specific emissivity is defined as
\begin{equation}
  \epsilon_{\nu} = f_{\ast} \overline{\rho}_\mathrm{b,0} \frac{df_{\rm{coll}}}{dt} l_{\nu},
\end{equation}
where $f_{\ast}$ is the star formation efficiency, $\overline{\rho}_{b,0}$ is the mean baryon density, $f_{\rm{coll}}$ is the fraction of matter in collapsed halos with virial temperatures in excess of $T_{\mathrm{min}}$, and $l_{\nu}$ encodes the efficiency of photon production (per stellar baryon) as a function of photon frequency. Note that the product $f_{\ast} \overline{\rho}_\mathrm{b,0} df_{\rm{coll}}/dt$ is equivalent to the star formation rate density (SFRD) of the Universe in this model.

\begin{figure*}[t]
\centering
\includegraphics[scale=0.75]{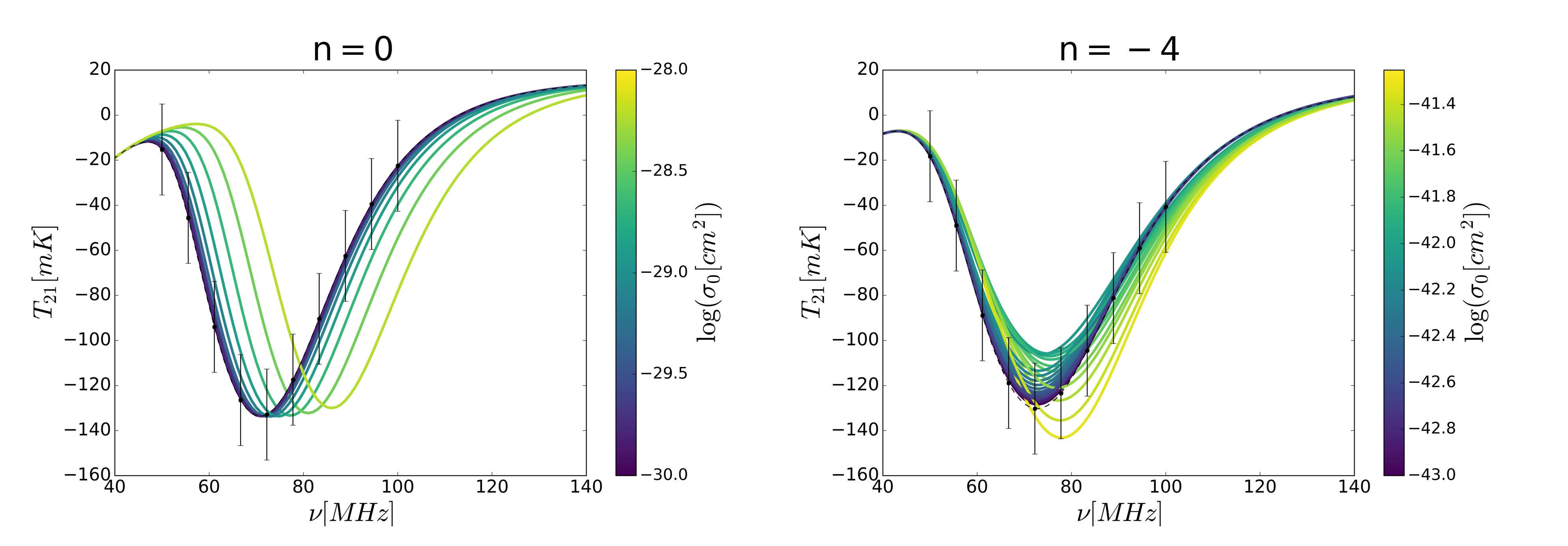}
\caption{The Global 21-cm signal as a function of frequency in IDM models with $n=0$ (left-hand panel) and $n=-4$ (right-hand panel), for DM particle mass of $1$ GeV, are shown. The error bars represents the rms noise of the EDGES experiment. The color scheme captures the change in the signal as a function of the interaction cross section.}
\label{plt:signals_1Gev_global}
\end{figure*}

We treat the Lyman-$\alpha$, Lyman-continuum, and X-ray bands separately, assigning each its own distinct efficiency factor $l_{\nu}$. For the Lyman-$\alpha$ photons, we set $l_{\nu} = N_{\rm{lw}}$, where $N_{\rm{lw}}$ is the number of photons emitted per stellar baryon. For ionizing photons, $l_{\nu} = f_{\mathrm{esc}} N_{\mathrm{ion}}$. Here, we account for the possibility that only a fraction, $f_{\mathrm{esc}}$, of the ionizing photons emitted by stars, $N_{\mathrm{ion}}$, actually escape their host galaxies. Finally, for X-rays, $l_{\nu}$ is linked to the efficiency of X-ray production in nearby star-forming galaxies. The soft X-ray luminosity $L_\mathrm{X}$ follows the relation $L_\mathrm{X} \simeq 2.6 \times 10^{39} \mathrm{erg} \ \rm{s}^{-1} (M_{\odot} / \rm{yr})^{-1} \left(\mathrm{SFR} / (M_{\odot} / \rm{yr})\right)$, where SFR is the star formation rate~\cite{mineo2012x}. We introduce a free parameter $f_\mathrm{X}$ that scales this relation, i.e., $f_\mathrm{X} = 1$ indicates the local $L_\mathrm{X}/\rm{SFR}$ relation, while $f_\mathrm{X} > 1$ indicates more efficient X-ray production (per unit star formation) than $z \sim 0$ galaxies. Note that $N_{\mathrm{lw}}$, the product $f_{\mathrm{esc}} N_{\mathrm{ion}}$, and $\mathrm{f}_\mathrm{X}$ are all degenerate with $f_{\ast}$. Consequently, parameter forecasts are presented in terms of the product of $f_{\ast}$ and these parameters, and $f_{\ast}$ is not independently varied in this work. It is possible to break this degeneracy by independently constraining $f_{\ast}$ via high-$z$ galaxy luminosity functions~\cite{mirocha2017}, and performing a joint LF/21-cm likelihood analysis~\cite{hibbard2022constraining, dorigojones2023}. However, we defer an exploration of this approach to future work. 

With the UV and X-ray emissivity in hand, one can determine the Lyman-$\alpha$ intensity as a function of redshift, as well as the ionization and heating rates of the intergalactic medium (IGM). These calculations are performed within a two-zone model of the IGM, where the fully-ionized regions and "bulk IGM" outside ionized bubbles are modeled separately. As a result, the mean ionized fraction $\bar{x}_i$ reflects both the volume filling factor of ionized gas, typically denoted as $Q$, and the ionized fraction of the bulk neutral IGM, $x_e$, such that $\bar{x}_i = Q + x_e (1 - Q)$. The spin temperature represents the temperature of the largely neutral bulk IGM. For further details, see \cite{mirocha2014}.

We present the 21-cm Global signals computed within the IDM cosmology for various cross section values and $m_\mathrm{\chi}=1$ GeV in Fig.~\ref{plt:signals_1Gev_global}, along with the error bars for an EDGES-like experiment. The cosmological and astrophysical parameters are kept fixed at their fiducial values for this plot. The left-hand panel corresponds to the $n=0$ (velocity-independent cross section) model, while the right-hand panel represents the $n=-4$ (Coulomb-like interaction) model. From the left-hand panel, it is evident that for the velocity-independent cross section model, the signals show a smooth shift of the troughs towards higher frequencies as the cross section increases. However, in case of Coulomb-like interaction, the signal exhibits a non-monotonic behavior in the position of the troughs with increasing cross sections. This difference in the behavior of the 21-cm signal with cross section for two IDM models can be attributed to the competition of two effects: the dissipation of the relative bulk velocity of DM and baryons and the impact of interaction cross section on the efficiency of baryon cooling.

\section{Methods}\label{sec:methods_global}
We employ Fisher forecasting to perform statistical inference and determine the sensitivity of 21-cm global signal observations for future experiments. We first provide a concise overview of the Fisher forecasting formalism tailored to 21-cm observations. Subsequently, we discuss the input covariance noise models adopted in this analysis.

\subsection{Fisher Forecasting}
Fisher forecasting formalism is a statistical framework commonly used in cosmology to predict and quantify uncertainties in measurements of cosmological parameters by a given experiment or observation \cite{fisher1935logic}. It has been widely applied in the context of CMB and large-scale structure (LSS) observations~\cite{ade2019simons, coe2009fisher}. In this paper, we adopt this formalism to forecast the sensitivity of various future Global 21-cm experiments and predict constraints on both IDM model parameters and associated astrophysical parameters. The Fisher formalism estimates the sensitivity of these measurements by analyzing the curvature of the likelihood surface in the underlying parameter space. Such analyses can provide valuable insights into the potential scientific yield of these future experiments and help optimize the experimental design. 
The Fisher matrix is defined as the negative expected value of the second derivative of the log-likelihood function with respect to the underlying model parameters. Mathematically, it is expressed as:
\begin{eqnarray}
    F_\mathrm{ij} = - \left< \frac{\partial^2  \ln \mathcal{L} }{ \partial \theta_\mathrm{i}\partial \theta_\mathrm{j} } \right>\,,
\end{eqnarray}
where $F_\mathrm{ij}$ are the $(\mathrm{i}, \mathrm{j})$-th elements of the Fisher matrix, $\theta_\mathrm{i}$ are the model parameters, $\ln \mathcal{L}$ represents the log-likelihood function, and $<>$ denotes the ensemble average over different data realizations.

The Fisher matrix is often called the Fisher Information Matrix (FIM), and it is related to the covariance matrix ($C$) of the parameter estimates as follows:
\begin{eqnarray}
	C_\mathrm{ij} = F^{-1}_\mathrm{ij}\, ,
\end{eqnarray}
where $C_\mathrm{ij}$ are the $(\mathrm{i}, \mathrm{j})$-th elements of the covariance matrix and $F^{-1}_\mathrm{ij}$ are the $(\mathrm{i}, \mathrm{j})$-th element of the inverse of Fisher matrix. Once the Fisher matrix is calculated, one can compute the uncertainties (standard deviations) of individual parameters $\theta_\mathrm{i}$ as follows:
\begin{eqnarray}
	\sigma(\theta_\mathrm{i}) = \sqrt{F^{-1}_\mathrm{ii}}\,,
\end{eqnarray}
where $\sigma(\theta_\mathrm{i})$ is the uncertainty in measurement of parameter $\theta_\mathrm{i}$. The off-diagonal elements of the covariance matrix represent the correlations between different parameters. A positive value indicates a positive correlation, while a negative value indicates a negative correlation. The correlation coefficient ($\rho_\mathrm{ij}$) between the two parameters $\theta_\mathrm{i}$ and $\theta_\mathrm{i}$ can be defined as
\begin{eqnarray}
	\rho_\mathrm{ij} = \frac{C_\mathrm{ij}}{\sigma(\theta_\mathrm{i}) \sigma(\theta_\mathrm{j})}\,.
\end{eqnarray}
To forecast the parameter uncertainties for a given experiment or survey, one can compute the Fisher matrix elements based on the experimental setup.
Inverting the Fisher matrix then yields the covariance matrix and associated parameter uncertainties. For this study, we assume that the likelihood function follows a multivariate Gaussian distribution which allows for a quadratic approximation of the likelihood near its maximum, thereby simplifying the calculation of the Fisher matrix.
\begin{eqnarray}
	\mathcal{L}(\boldsymbol{\mu}|\boldsymbol{X})=\frac{\exp \Bigl(- \frac{1}{2}(\boldsymbol{X} - \boldsymbol{\mu})^{\mathrm{T}} \boldsymbol{\Sigma}^{-1} (\boldsymbol{X}-\boldsymbol{\mu}) \Bigl)}{\sqrt{(2\pi)^\mathrm{k} |\boldsymbol{\Sigma}|}}\,,
\end{eqnarray}
where $\boldsymbol{X}$ is the observed $k$-dimensional column vector, $\boldsymbol{\mu}$ is the model-predicted $k$-dimensional column vector, $\boldsymbol{\Sigma}$ is the symmetric input covariance matrix, and $|\boldsymbol{\Sigma}|$ is the determinant of the input covariance matrix.

In the case of 21-cm cosmology, the log-likelihood function is
\begin{eqnarray}
	\ln \mathcal{L}(\boldsymbol{\theta}|\boldsymbol{X}) \propto - \frac{1}{2} \Bigl( (\boldsymbol{X} - T(\boldsymbol{\theta}))^{T} \boldsymbol{\Sigma}^{-1} (\boldsymbol{X}-T(\boldsymbol{\theta})) \Bigl)\,,
\end{eqnarray}
where $\boldsymbol{\theta}$ is the parameter column vector and $T(\boldsymbol{\theta})$ is the temperature of the global 21-cm signal generated by \textsc{ares} code for $\boldsymbol{\theta}$ as its input parameters. In the following subsection, we will elaborate more on the models used for the input covariance matrix.

\begin{table*}[t] 
\centering 
\resizebox{\textwidth}{!}{
\begin{tabular}{|c|c|c|c|c|c|c|c|}  
\hline  
\hline 
Experiment & $\nu_\mathrm{min},\nu_\mathrm{max}$ (MHz) & $N_\mathrm{ch}$ & $\Delta\nu$ (kHz) & $\tau_\mathrm{obs}$ (h) & $T_\mathrm{rec}$ (K) & $\sigma_\mathrm{RMS} \ (\mathrm{mK})$ & Input Noise\\
\hline  
EDGES-like & 50-100 & 10 & 390.625 & 107 & 300 & 20.2 & RMS\\
SARAS-like & 55-85 & 10 & 61 & 100 & 300 & 23.4 & RMS\\
Future1 & 50-100 & 10 & 390.625 & 200 & 300 & - & Radiometer\\
Future2 & 50-100 & 100 & 390.625 & 1000 & 300 & - & Radiometer\\
\hline  
\end{tabular}}
\caption{ Comparison of four experimental setups, outlining their key parameters such as minimum and maximum frequencies probed ($\nu_\mathrm{min},\nu_\mathrm{max}$), number of independent frequency channels ($N_\mathrm{ch}$), native (not smoothed) frequency bins ($\Delta\nu$), observation times ($\tau_\mathrm{obs}$), receiver temperatures ($T_\mathrm{rec}$), and RMS value of the smoothed noise ($\sigma_\mathrm{RMS}$).}
\label{tab:experiment_specs}
\end{table*}

\subsection{Noise Models}\label{subsec:noisemodels}
A crucial step in Fisher forecasting is choosing an appropriate input noise model. In this analysis, we adopt two noise models. The first model is based on the root mean square (RMS) noise, which is computed by fitting both the foreground and global signal models to the actual data obtained by a given experiment. The second noise model uses radiometer noise, which is the natural noise of a radio receiver. In this study, we consider four experimental setups as mentioned in Table~\ref{tab:experiment_specs} and plotted in Fig.~\ref{plt:noises}. We apply RMS noise for EDGES-like and SARAS-like experiments, while radiometer noise for Future1 and Future2 experiments.

Ref.~\cite{hills2018concerns} demonstrated that only a limited number of frequency channels from the EDGES experiment are effectively independent. The reason is the presence of structures with a period of less than $\sim10$ MHz in the residual plot of EDGES data after subtracting the foreground and the template of the 21-cm signal. Using the same method employed in Ref.~\cite{hills2018concerns} to estimate spectral structures, we identified features with a characteristic scale of less than $\sim6$ MHz in the SARAS data. Based on this, we assume ten independent data points for both EDGES-like and SARAS-like experiments. Accordingly, we model the input covariance matrix as diagonal, implying no correlations between different frequency channels.

\subsubsection{RMS (White) Noise}
As mentioned above, the RMS values for the noise level for each experiment depend on the smoothing frequency. Specifically, the RMS values are $20.2$ mK and $23.4$ mK at smoothing frequencies of $3.12$ MHz and $2.8$ MHz for the EDGES~\cite{edges_nature} and SARAS3~\cite{singh2022detection} experiments, respectively. These values are consistent with the data presented in Fig.~$5$b of Ref.~\cite{singh2022detection}. The higher noise levels in the SARAS3 experiment can primarily be attributed to its lower integration time, though improvements are expected in the future.  We use these RMS noises in our analysis for EDGES-like and SARAS-like experimental setups (see the first two rows in Table~\ref{tab:experiment_specs} for details).

\subsubsection{Radiometer Noise}
The noise of an ideal radiometer with an integration time of $\tau$ and channel width $\Delta\nu$ with a system temperature $T_\mathrm{sys}$ is given by \cite{rohlfs2013tools, furlanetto2006cosmology}
\begin{eqnarray}\label{eq:radio_noise}
\sigma_\mathrm{T} = \frac{T_\mathrm{sys}}{\sqrt{\Delta\nu \ \tau}} \,,
\end{eqnarray}
where $T_\mathrm{sys} = T_\mathrm{sky} + T_\mathrm{receiver}$ with  $T_\mathrm{sky}$ and $T_\mathrm{receiver}$ being the temperatures of sky and receiver, respectively. We calculate the sky temperature using the $408$ MHz Haslam all-sky map~\cite{haslam1982408}, which is defined as
\begin{eqnarray}
T_\mathrm{sky} = T_{408} \left(\frac{\nu}{\nu_{408}} \right)^{\beta}\ ,
\end{eqnarray}
where we fix $\beta=-2.56$ and $T_{408}=22.7$ K. Following the calculations presented in Ref.~\cite{sims2020testing}, we consider $T_\mathrm{receiver} \approx 300$ K, $\tau = 107$ h, $\Delta\nu = 390.625$ kHz for the EDGES-like experiment. Since Future1 and Future2 experimental setups are enhanced versions of the EDGES-like experiment, featuring longer integration times and additional frequency channels, we also apply the same specifications to these scenarios.

\begin{figure}[t]
\centering
\includegraphics[scale=0.38]{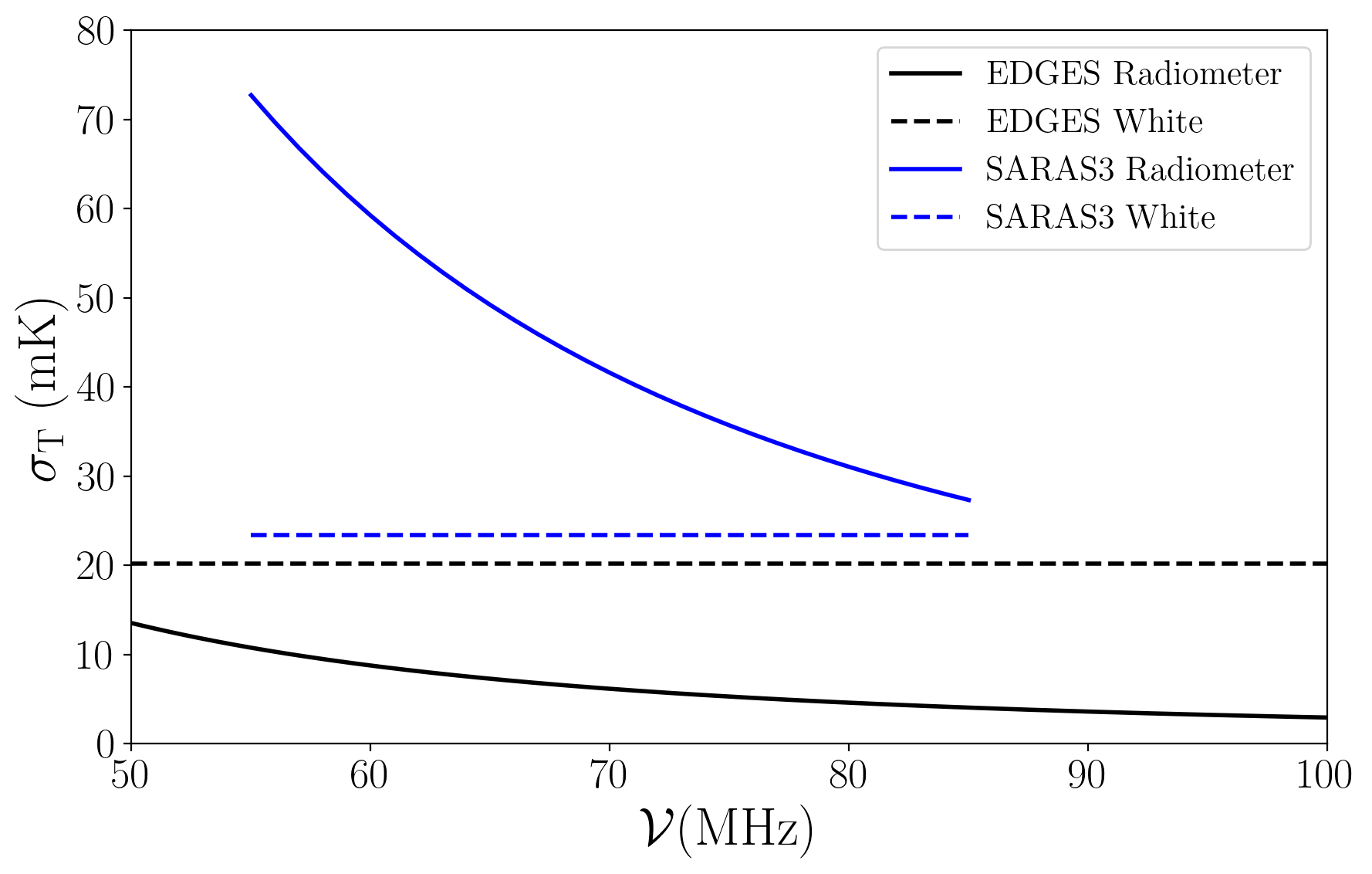}
\caption{The Radiometer and RMS (white) noise for EDGES and SARAS3 experiments are shown in solid and dashed lines, respectively. Radiometer noise levels are calculated using the specifications listed in Table~\ref{tab:experiment_specs}, while the white noise levels are calculated using the smoothed noise reported by the respective experiment~\cite{edges_nature, singh2022detection}.}
\label{plt:noises}
\end{figure}

Fig.~\ref{plt:noises} shows the input noise as a of function frequency for both noise models. It is evident from this figure that the white noise levels in a SARAS-like at present stage are comparable to that in an EDGES-like experiment. According to Eq.~\ref{eq:radio_noise}, the critical parameter to reduce the value of radiometer noise is the product of the channel width and the integration time. In other words, reducing the channel width requires increasing the integration time to keep a constant noise level, assuming all other parameters remain unchanged. This suggests that a higher integration time is needed for SARAS-like experiments to maintain the same level of radiometer noise compared to an EDGES-like experiment. Hence, for our analysis, we consider only the noise levels associated with EDGES for our futuristic scenarios. However, the pipeline presented here can be easily adapted to accommodate different noise levels for future and ongoing experiments.

\begin{table}[t] 
\centering  
\begin{tabular}{|c|c|}  
\hline  
Parameter & Fiducial Value\\
\hline  
$f_\mathrm{*}$ & 0.01 (fixed)\\
$f_\mathrm{esc}$ & 0.1 \\
$f_\mathrm{X}$ & 1.0\\
$T_\mathrm{min}$ & 500\\
$N_\mathrm{lw}$ & 9690\\
$m_\mathrm{\chi}$ & 100 kev-1 Tev\\
$\sigma_\mathrm{0}$ & 0.0 (CDM)\\
\hline  
\end{tabular}
\caption{Fiducial value of IDM and astrophysical parameters used in the Fisher analysis.}
\label{tab:fid_vals_global}
\end{table}

\section{Results}\label{sec:results_global}
Throughout this paper, we assume $\Lambda$CDM cosmology as our fiducial model. Given the discrepancies between the results reported by the EDGES and SARAS3 experiments, we do not attempt to analyze their data sets. Instead, we use the two experiments as realistic experimental test cases to assess the sensitivity of the global signal to DM physics. Our approach is to generate a mock realization of the global 21-cm signal within the $\Lambda$CDM cosmology and assess the sensitivity of each experimental configuration, assuming noise levels similar to those of EDGES and SARAS3. 

We assume that the observed signal is consistent with the standard $\Lambda$CDM cosmology predictions and infer the upper bound on DM-baryon interaction cross section in the case of a null detection. However, the pipeline presented in this work is designed to be flexible and can be easily extended to incorporate different noise and fiducial models. This adaptability ensures that the framework remains applicable for future explorations of the data as noise levels decrease in future observations.

As explained in Section~\ref{sec:21-cm_model}, we utilize the modified version of three codes, \textsc{class}, \textsc{galacticus}, and \textsc{ares}, as a single pipeline to generate the global 21-cm signals within the IDM cosmology. We then use these generated signals to perform the Fisher forecast analysis with two different input covariance noise models to estimate the sensitivities of four experimental scenarios. An example of such global 21-cm signals is presented in Fig.~\ref{plt:signals_1Gev_global}.

\begin{figure*}
\centering
\includegraphics[scale=0.55]{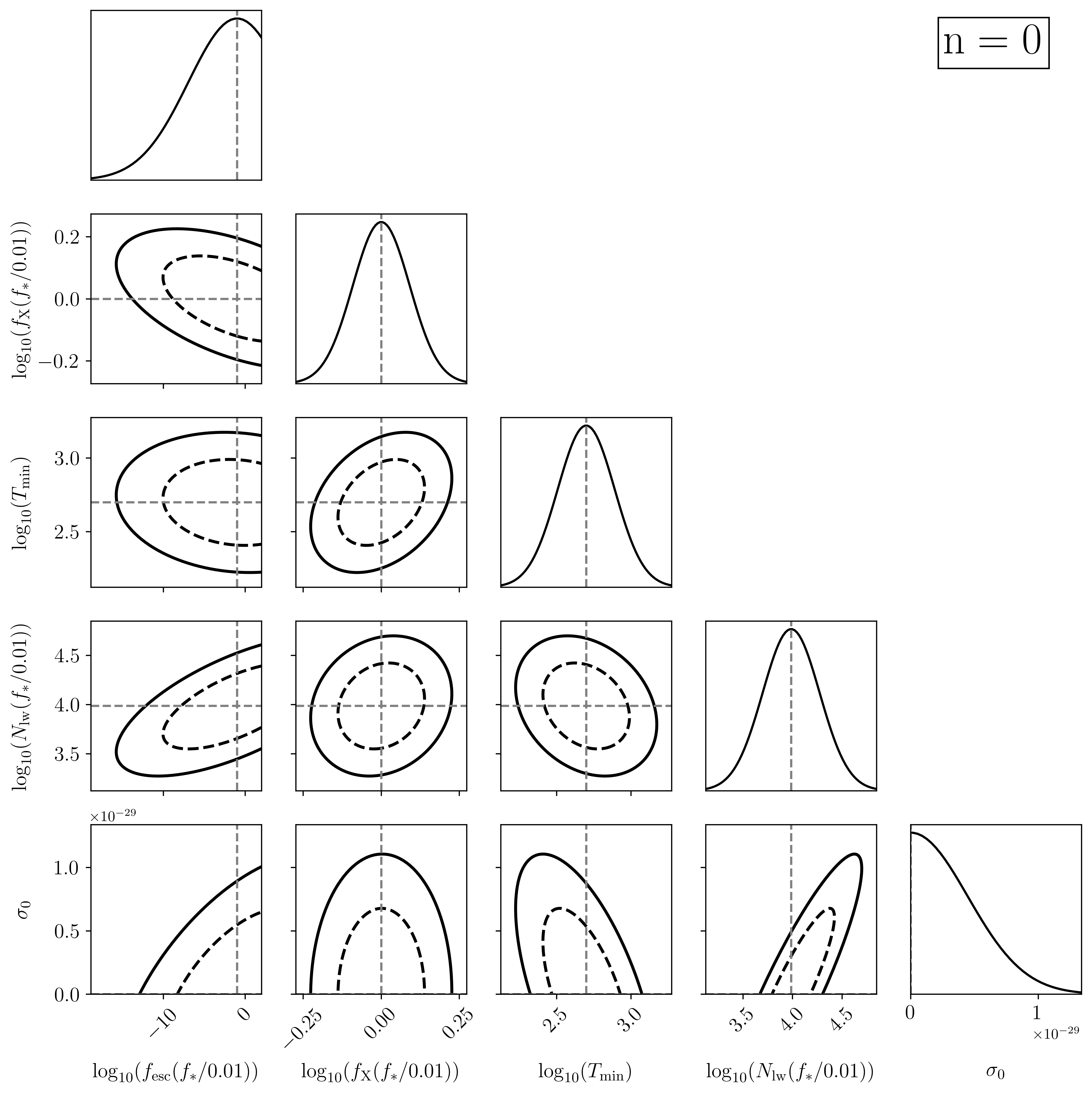}
\caption{Forecasts on astrophysical parameters and the cross section of interactions for the velocity-independent cross section IDM model with a fixed DM mass of $10$ MeV are presented. These forecasts are obtained for an EDGES-like experiment, using the RMS noise model for the input covariance matrix. The two-dimensional contours represent the $68\%$ and $95\%$ confidence regions of the posterior probability distribution, shown by the dotted and solid black curves, respectively, while the top panel of each represents the marginalized posterior distributions for the corresponding parameters. The gray lines correspond to the fiducial values of each parameter.}
\label{plt:n0_tri_global}
\end{figure*}

\begin{figure*}
\centering
\includegraphics[scale=0.55]{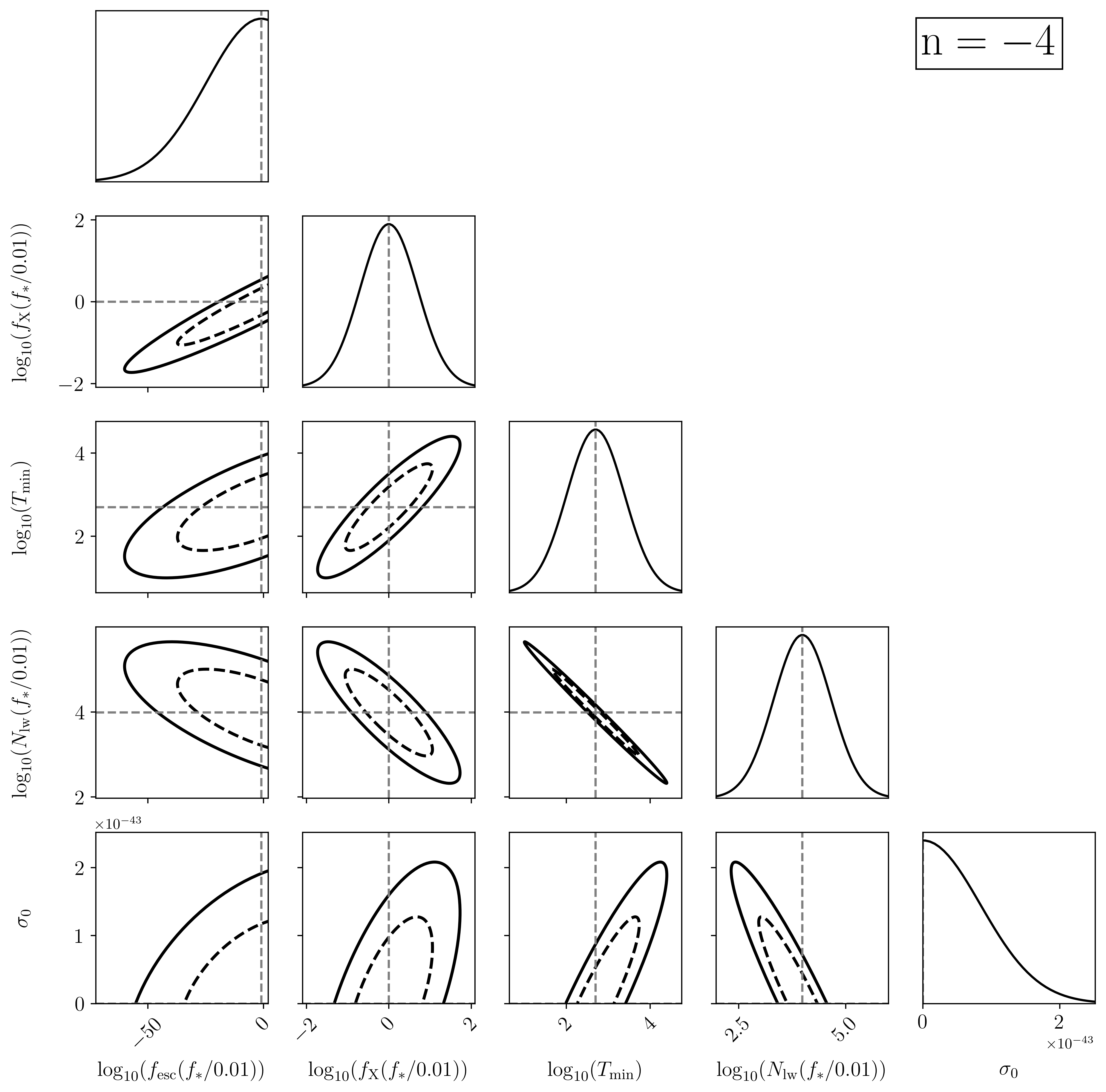}
\caption{Similar to the triangle plot in Fig.~\ref{plt:n0_tri_global} for $n = 0$, we present the forecasts on astrophysical parameters and $\mathrm{\sigma}_{0}$ for the Coulomb-like IDM model with a fixed DM mass of $10$ MeV. These forecasts are obtained for an EDGES-like experiment, using the RMS noise model for the input covariance matrix. The two-dimensional contours represent the $68\%$ and $95\%$ confidence regions of the posterior probability distribution, shown by the dotted and solid black curves, respectively, while the top panel of each column represents the marginalized posterior distributions for the corresponding parameters. The gray lines correspond to the fiducial values of each parameter.}
\label{plt:n-4_tri_global}
\end{figure*}

We perform a multi-parameter Fisher forecasting analysis with the following set of astrophysical and IDM parameters: $ \boldsymbol{\theta} =[f_\mathrm{esc}, f_\mathrm{X}, T_\mathrm{min}, N_\mathrm{lw}, \sigma_\mathrm{0}]$. The fiducial values for each parameter are summarized in Table~\ref{tab:fid_vals_global}, and additional details about these parameters are provided in Section~\ref{sec:21-cm_model}. It is important to note that we have kept $f_\mathrm{\ast}$ fixed, as it is highly degenerate with nearly all other astrophysical parameters considered in this analysis. After calculating the Fisher matrix, we use \textsc{fishchips} \footnote[4]{\url{https://github.com/xzackli/fishchips-public}} \cite{li2018disentangling} to visualize the 2-D and 1-D marginalized posteriors. These results for IDM models with $n = 0$ and $n = -4$ are shown in Figs.~\ref{plt:n0_tri_global} and \ref{plt:n-4_tri_global}, respectively.

\begin{figure*}
\includegraphics[scale=0.55]{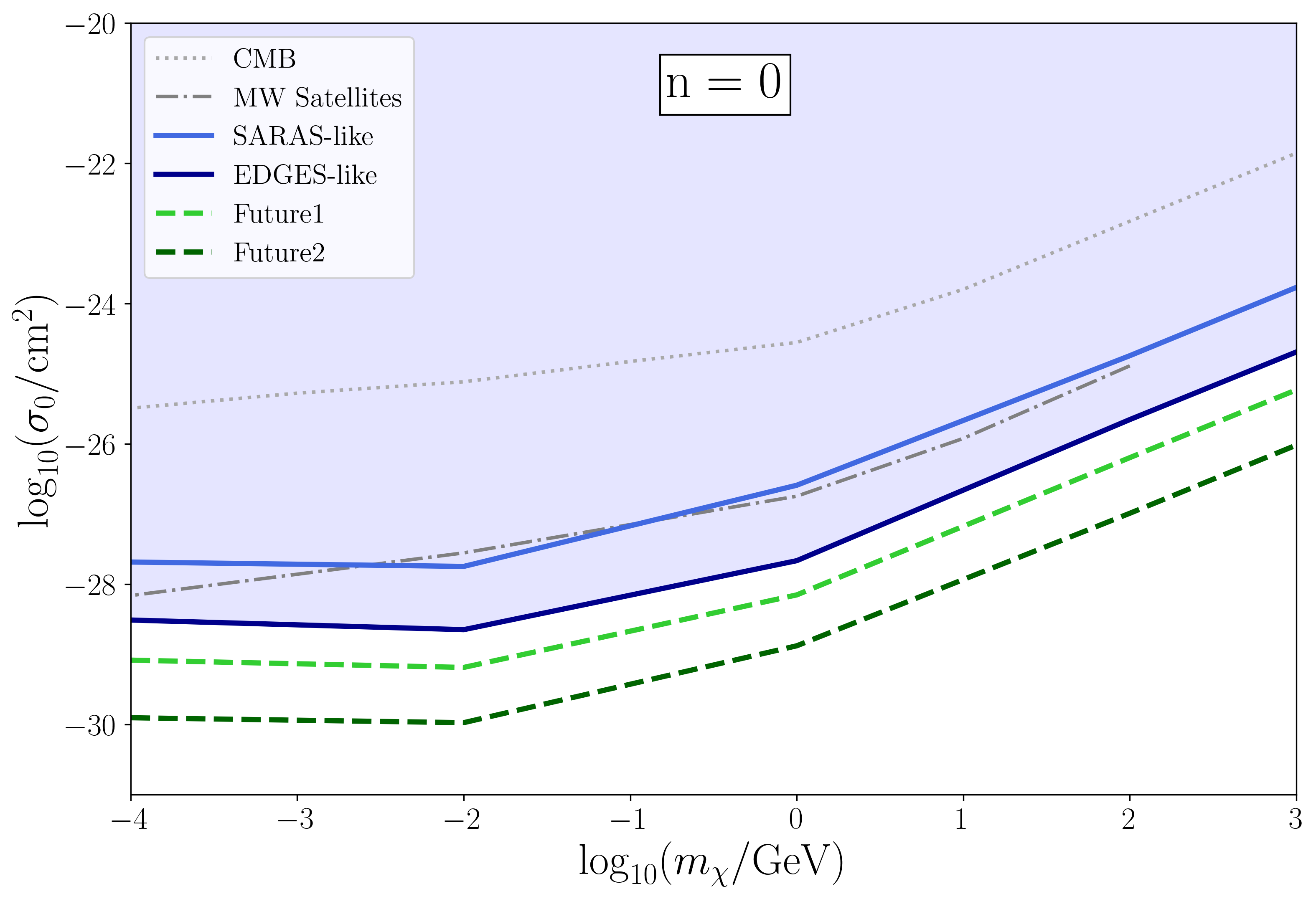}
\caption{The 95\% confidence level upper limit forecasts on $\log_{10}(\sigma_{0}/\mathrm{cm}^2)$ as a function of IDM mass are shown for a velocity-independent cross section IDM model ($n=0$) and four different experimental scenarios. The previous bounds from CMB~\cite{nguyen2021observational} and Milky Way satellite abundance~\cite{maamari2021bounds} are also plotted for the comparison. While a SARAS-like experiment shows results comparable to the current best bound, the other three experimental scenarios predict stronger forecasts.}
\label{plt:n0_main_plot_global}
\end{figure*}

\begin{figure*}
\includegraphics[scale=0.55]{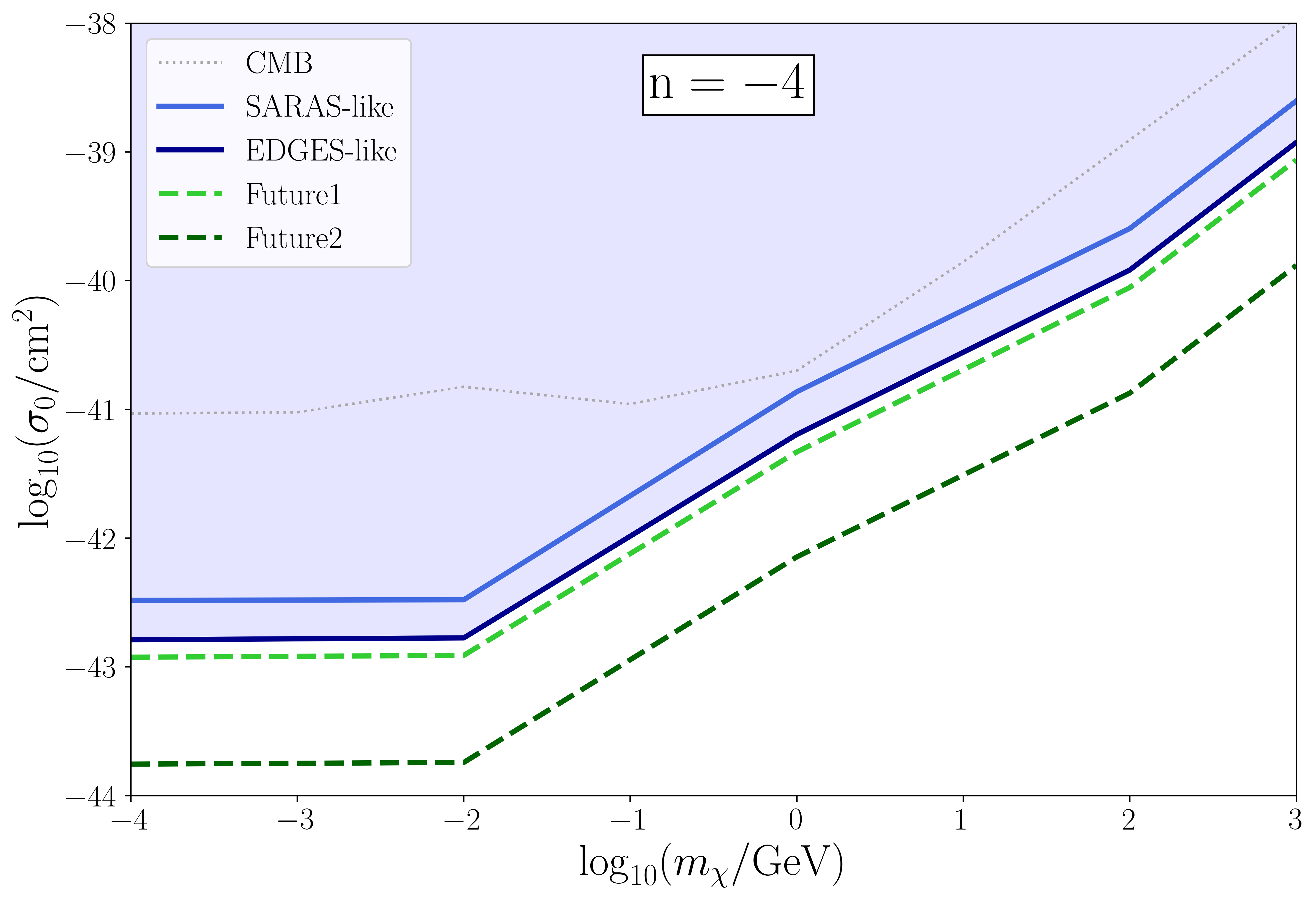}
\caption{The 95\% confidence level upper limit forecasts on $\log_{10}(\sigma_{0}/\mathrm{cm}^2)$ as a function of IDM mass are shown for Coulomb-like IDM model ($n=-4$) and four different experimental scenarios (the shaded region, or area above a given sensitivity curve, would be excluded by a non-detection of a signal with a given experiment). The previous bounds from CMB \cite{nguyen2021observational} are also plotted for comparison. All four experimental scenarios predict stronger forecasts compared to the current CMB bounds.}
\label{plt:n-4_main_plot_global}
\end{figure*}

Fig.~\ref{plt:n0_tri_global} presents the two-dimensional $68\%$, $95\%$ confidence contours for all parameters of interest, showing the correlation between different astrophysical parameters and the interaction cross section for the $n=0$ IDM model. The top panel in each column of the figure shows the marginalized posterior distributions for the corresponding parameters. It is evident from the plot that $\sigma_\mathrm{0}$ has no correlation with $f_\mathrm{X}$ and features an anti-correlation with $T_\mathrm{min}$. On the one hand, this suggests that the interaction cross section can be reconstructed with little bias from astrophysical uncertainty related to the X-ray luminosity of early galaxies. On the other hand, the increase in the minimum virial temperature delays the formation of the first galaxies, causing a delay in the heating and ionization of the intergalactic medium, which shifts the trough of the 21-cm signal to higher frequencies; a velocity-independent interaction between DM and baryons affects the 21-cm signal similarly. A minimum virial temperature, related to the minimum mass of halos forming galaxies, is therefore a major source of uncertainty in DM interaction inference with the global 21-cm signal. We further find a significant positive correlation between $\sigma_\mathrm{0}$ and $N_\mathrm{lw}$, suggesting that an increase in the interaction cross section requires a higher number of Lyman-Werner photons to compensate for the signal shift caused by the enhanced interaction rate. The uncertainty in $N_\mathrm{lw}$ is therefore likewise of concern to DM inference with the global signal. Finally, the broad uncertainties in the forecasts for $f_\mathrm{esc}$ indicate that the global 21-cm signal is largely insensitive to this parameter within the frequency range considered across the four scenarios. This is because $f_\mathrm{esc}$ primarily affects the reionization process, which is expected to occur mostly at $z \lesssim 10$ based on other observational constraints, outside the most sensitive redshift range probed here.

Fig.~\ref{plt:n-4_tri_global} presents the two-dimensional $68\%$, $95\%$ confidence contours, showing the forecasts for different astrophysical parameters and the cross section of interaction for the IDM model with Coulomb-like interaction ($n=-4$). We find a positive correlation between $\sigma_\mathrm{0}$ and $f_\mathrm{X}$, as both parameters tend to reduce the depth of the global signal and move it to lower frequencies (for low enough cross sections). Furthermore, increasing the cross section delays the structure formation, which is degenerate with a decrease in the number of Lyman-Werner photons ($N_\mathrm{lw}$). However, if baryons cool through interactions with DM, a lower virial temperature, an increase in the number of Lyman-Werner photons, and an increase in X-ray efficiency is needed to preserve the amplitude of the signal. We also note a strong anti-correlation between $T_\mathrm{min}$ and $N_\mathrm{lw}$, indicating that the two parameters are difficult to constrain independently with global signal data alone. Similar to the $n = 0$ case, $f_\mathrm{esc}$ is not highly correlated with any other parameters, as its impact on the signal becomes significant only at high frequencies. 

Fig.~\ref{plt:n0_main_plot_global} is derived by repeating the analysis represented by Fig.~\ref{plt:n0_tri_global}, for a range of DM particle masses, and marginalizing over all astrophysical parameters. This figure shows a forecast of the $95\%$ confidence level upper bound on the interaction cross section between DM and baryons for the $n = 0$ model, as a function of the DM particle mass. Forecasts are computed for four different experimental scenarios, as listed in Table~\ref{tab:experiment_specs} and discussed in Subsection~\ref{subsec:noisemodels}. The region of the parameter space above each curve would be excluded at the $95\%$ confidence level by a given experiment. For example, the shaded region in the plot corresponds to the region excluded by the EDGES-like configuration. The current bounds from CMB Planck data (dotted grey)~\cite{nguyen2021observational} and from the Milky Way satellite population measurements (dot-dashed grey)~\cite{maamari2021bounds} are also shown for comparison. The light blue and dark blue curves correspond to constraints from SARAS-like and EDGES-like scenarios, respectively. Notably, the SARAS-like scenario, which has the lowest sensitivity among the four cases explored in this work, can achieve a comparable upper bound to the best existing bounds derived using the Milky Way satellite data~\cite{maamari2021bounds}. The light and dark dotted green curves show the forecasts for Future1 and Future2 experimental setups, respectively. These forecasts illustrate the improvement in sensitivity with increasing observation time and more frequency channels.

Fig.~\ref{plt:n-4_main_plot_global} similarly shows the 95\% forecasts on the cross section of interaction between DM and baryons for the $n=-4$ model, for a range of DM particle masses. Similar to the forecast plot for the $n=0$ model, these forecasts are computed for four different experimental scenarios. The previous bounds from CMB (dotted grey) \cite{nguyen2021observational} are also plotted for comparison. The light and dark blue curves correspond to the constraints for SARAS-like and EDGES-like scenarios. To emphasize the forecasts for the EDGES-like scenario, the excluded parameter space is shaded in blue. It can also be seen in Fig.~\ref{plt:n-4_main_plot_global} that the global signal is highly sensitive to the $n=-4$ model (see also Fig.~\ref{plt:signals_1Gev_global}). Consequently, the SARAS-like experiment can in principle surpass the sensitivity of the CMB data to these interaction scenarios. This improvement is particularly pronounced for lower masses, where the forecasted upper bound on the interaction cross section is more than an order of magnitude lower than the current observational bounds. Additionally, we show the forecasts for two futuristic scenarios in the light-dotted and dark-dotted green curves to highlight the potential of future 21-cm measurements in exploring DM interactions. The corresponding projected 95\% confidence level upper limits are also presented in Table~\ref{tab:final_n0,-4_global}. We can also translate the constraints on $\sigma_0$ to the physical cross-section by multiplying with the factor of $\left(v_{\mathrm{rel}}/c\right)^{n}$, where $c$ is the speed of light. For $n = 0$, $\sigma_{0}$ is identical to the physical cross section, while for Coulomb-like interactions they differ by a factor of $\left(v_{\mathrm{rel}}/c\right)^{-4}$.   For completeness, we provide the multiplicative factors required to convert $\sigma_{0}$ to the physical cross section $\sigma$ at two representative epochs: $z = 0$ and $z = 18$. The latter corresponds to the cosmic epoch near the absorption trough of the global 21-cm signal reported by the EDGES collaboration, while $z = 0$ represents the present day for reference. At redshifts $z=0$ and $z=18$, the relative bulk velocities are $v_{\mathrm{rel}} = 29.97~\mathrm{m/s}$ and $v_{\mathrm{rel}} = 569.4~\mathrm{m/s}$, respectively,  yielding multiplicative conversion factors of $1.00\times10^{28}$ and $7.68\times10^{22}$, respectively.  The projected 95\% confidence level upper limits on the physical cross section for each experimental configuration can thus be obtained by applying these corresponding factors to the reported constraints on $\sigma_{0}$.

\section{Conclusions}\label{sec:conclusions_global}
In this study, we use the Fisher forecasting formalism to project the sensitivity of current and future global 21-cm signal experiments to detecting DM-baryon elastic scattering interactions. Scattering between DM and baryons alters structure formation and leads to a delay in the onset of the global absorption trough, and it also affects the thermal history of baryons. We model both effects in this work and assess their detectability, across several current and future experimental scenarios. We consider two interacting DM models: Coulomb-like interaction and velocity-independent interaction.

\begin{table}[t] 
\centering 
\begin{tabular}{|c|c|c|c|c|c|}  
\hline 
n & $M_\mathrm{\chi}$ & EDGES-like & SARAS-like & Future1 & Future2\\
\hline 
  & 100 kev & -28.51 & -27.68 & -29.90 & -29.08\\
  & 10 Mev  & -28.65 & -27.75 & -29.97 & -29.19\\
0 & 1 Gev   & -27.67 & -26.59 & -28.88 & -28.15\\
  & 100 Gev & -25.65 & -24.74 & -26.99 & -26.20\\
  & 1 Tev   & -24.69 & -23.77 & -26.02 & -25.23\\
\hline
  &   100 kev & -42.79 & -42.48& -42.92& -43.76\\
  & 10 Mev  & -42.77 & -42.48& -42.91& -43.74\\
-4 & 1 Gev   & -41.19 & -40.86& -41.33& -42.15\\
  & 100 Gev & -39.92 & -39.59& -40.05& -40.87\\
  & 1 Tev   & -38.93 & -38.60& -39.06& -39.88\\
\hline 
\end{tabular}
\caption{The 95\% confidence level upper limit forecasts on $\mathrm{\log}_{10}(\sigma_{0}/\mathrm{cm}^2)$, the coefficient of the momentum-transfer cross section of DM–baryon scattering for the velocity independent ($n=0$) and Coulomb-like ($n=-4$) models, for different IDM masses and four experimental setups listed in Table~\ref{tab:experiment_specs} are presented.}
\label{tab:final_n0,-4_global}
\end{table}

For a velocity-independent elastic scattering, we find that an experiment with sensitivity comparable to EDGES can already provide stronger constraints on the cross section than the current best bounds derived from the Milky Way substructure, while a SARAS-like experiment has a sensitivity comparable to the existing bounds. However, in the $n=-4$ case, even the current experimental configurations can in principle yield stronger upper limit on the interaction cross section, should they measure a global signal consistent with standard $\Lambda$CDM cosmology. This is because the signal for $n=-4$ is more sensitive to the cross section compared to the $n=0$ model. Additionally, the 21-cm global signal shows more constraining power in the lower mass range in all four scenarios.

We further quantify in detail the degeneracy between key astrophysical parameters that affect the global 21-cm signal, and parameters that describe new DM interaction physics, exploring the significance of understanding the uncertainties inherent in global 21-cm signal measurements to DM inference with 21-cm data. We find that the minimum virial temperature is highly degenerate with DM-baryon interaction cross section, presenting a positive correlation in the case of $n=0$, and a negative correlation in the case of $n=-4$. We similarly find that the overall number of Lyman-Werner photons emitted by early galaxies is degenerate with the effects introduced by DM-baryon scattering. These are key astrophysical parameters whose accurate modeling or constraining through other observables presents key to DM inference with future global signal measurements. We further find that the inference of the interaction cross section for velocity-independent scattering is robust against uncertainties in X-ray luminosity modeling.

Overall, our results suggest that the global 21-cm signal can be a powerful probe for studying DM interactions. However, to achieve the full potential of the global signal measurements from the current and future experiments, it is necessary to understand the astrophysics of early galaxies, in particular their minimum virial temperature and processes that relate to the generation of Lyman-Werner photons, using complementary probes. Our initial exploration motivates a more detailed study of the complex parameter space that describes the global signal, and a pursuit of understanding the complementarity of different high-redshift probes of DM interactions. The insights gained from this study lay the groundwork for quantifying the power of 21-cm cosmology in new physics searches.

\begin{acknowledgments} 
A.R. acknowledges discussions with Ethan Nadler, and Andrew Benson. A.R. would like to thank Saurabh Singh for providing access to SARAS data. J.M. was supported by an appointment to the NASA Postdoctoral Program at the Jet Propulsion Laboratory/California Institute of Technology, administered by Oak Ridge Associated Universities under contract with NASA. V.G. and R.A. acknowledge the support from NASA through the Astrophysics Theory Program, Award Number 21-ATP21-0135. V.G. acknowledges the support from the National Science Foundation (NSF) CAREER Grant No. PHY-2239205, and from the Research Corporation for Science Advancement under the Cottrell Scholar Program.
\end{acknowledgments}

\bibliographystyle{JHEP}
\bibliography{bibliography}

@PREAMBLE{
 "\providecommand{\noopsort}[1]{}" 
 # "\providecommand{\singleletter}[1]{#1}%" 
}

@article{sims2020testing,
    author = {Sims, Peter H and Pober, Jonathan C},
    title = {Testing for calibration systematics in the EDGES low-band data using Bayesian model selection},
    journal = {Monthly Notices of the Royal Astronomical Society},
    volume = {492},
    number = {1},
    pages = {22-38},
    year = {2019},
    month = {12},
    issn = {0035-8711},
    doi = {10.1093/mnras/stz3388},
    url = {https://doi.org/10.48550/arXiv.1910.03165},
    eprint = {arXiv:1910.03165},
}

@article{singh2022detection,
  title={On the detection of a cosmic dawn signal in the radio background},
  author={Singh, Saurabh and Nambissan T, Jishnu and Subrahmanyan, Ravi and Udaya Shankar, N and Girish, BS and Raghunathan, A and Somashekar, R and Srivani, KS and Sathyanarayana Rao, Mayuri},
  journal={Nature Astronomy},
  volume={6},
  number={5},
  pages={607--617},
  year={2022},
  publisher={Nature Publishing Group UK London},
  doi = {https://doi.org/10.1038/s41550-022-01610-5},
  url = {https://doi.org/10.48550/arXiv.2112.06778},
  eprint = {arXiv:2112.06778}        
}

@article{dorigojones2023,
  title={Validating Posteriors Obtained by an Emulator When Jointly Fitting Mock Data of the Global 21 cm Signal and High-z Galaxy UV Luminosity Function},
  author={Jones, J Dorigo and Rapetti, D and Mirocha, J and Hibbard, JJ and Burns, JO and Bassett, N},
  journal={The Astrophysical Journal},
  volume={959},
  number={1},
  pages={49},
  year={2023},
  publisher={IOP Publishing},
  doi = {10.3847/1538-4357/ad003e},
  url = {https://doi.org/10.48550/arXiv.2310.02503},
  eprint = {arXiv:2310.02503} 
}

@article{mirocha2017,
    author = {Mirocha, Jordan and Furlanetto, Steven R. and Sun, Guochao},
    title = {The global 21-cm signal in the context of the high-z galaxy luminosity function},
    journal = {Monthly Notices of the Royal Astronomical Society},
    volume = {464},
    number = {2},
    pages = {1365-1379},
    year = {2016},
    month = {09},
    issn = {0035-8711},
    doi = {10.1093/mnras/stw2412},
    url = {https://doi.org/10.48550/arXiv.1607.00386},
    eprint = {arXiv:1607.00386},
}

@article{mirocha2014,
    author = {Mirocha, Jordan},
    title = {Decoding the X-ray properties of pre-reionization era sources},
    journal = {Monthly Notices of the Royal Astronomical Society},
    volume = {443},
    number = {2},
    pages = {1211-1223},
    year = {2014},
    month = {07},
    issn = {0035-8711},
    doi = {10.1093/mnras/stu1193},
    url = {https://doi.org/10.48550/arXiv.1406.4120},
    eprint = {arXiv:1406.4120},
}

@article{Zygelman2005,
  url = {https://ui.adsabs.harvard.edu/abs/2005ApJ...622.1356Z},
  title={Hyperfine level-changing collisions of hydrogen atoms and tomography of the dark age universe},
  author={Zygelman, B},
  journal={The Astrophysical Journal},
  volume={622},
  number={2},
  pages={1356},
  year={2005},
  publisher={IOP Publishing},
  doi={10.1086/427682}
}

@article{singh2017first,
  title={First results on the epoch of reionization from first light with SARAS 2},
  author={Singh, Saurabh and Subrahmanyan, Ravi and Shankar, N Udaya and Rao, Mayuri Sathyanarayana and Fialkov, Anastasia and Cohen, Aviad and Barkana, Rennan and Girish, BS and Raghunathan, A and Somashekar, R and others},
  journal={The Astrophysical Journal Letters},
  volume={845},
  number={2},
  pages={L12},
  year={2017},
  publisher={IOP Publishing},
  doi = {https://doi.org/10.3847/2041-8213/aa831b},
  url = {https://doi.org/10.48550/arXiv.1703.06647},
  eprint = {arXiv:1703.06647},
}

@article{edges_nature,
  title={An absorption profile centred at 78 megahertz in the sky-averaged spectrum},
  author={Bowman, Judd D and Rogers, Alan EE and Monsalve, Raul A and Mozdzen, Thomas J and Mahesh, Nivedita},
  journal={Nature},
  volume={555},
  number={7694},
  pages={67--70},
  year={2018},
  publisher={Nature Publishing Group UK London},
  doi={10.1038/nature25792},
  url = {https://doi.org/10.48550/arXiv.1810.05912},
  eprint = {arXiv:1810.05912},
}

@article{hills2018concerns,
  title={Concerns about modelling of the EDGES data},
  author={Hills, Richard and Kulkarni, Girish and Meerburg, P Daniel and Puchwein, Ewald},
  journal={Nature},
  volume={564},
  number={7736},
  pages={E32--E34},
  year={2018},
  publisher={Nature Publishing Group UK London},
  doi={https://doi.org/10.1038/s41586-018-0796-5},
  url = {https://doi.org/10.48550/arXiv.1805.01421},
  eprint = {arXiv:1805.01421},
}

@article{philip2019probing, 
  title={Probing radio intensity at high-z from marion: 2017 instrument},
  author={Philip, L and Abdurashidova, Z and Chiang, HC and Ghazi, N and Gumba, A and Heilgendorff, HM and J{\'a}uregui-Garc{\'\i}a, JM and Malepe, K and Nunhokee, CD and Peterson, J and others},
  journal={Journal of Astronomical Instrumentation},
  volume={8},
  number={02},
  pages={1950004},
  year={2019},
  publisher={World Scientific},
  doi={10.1142/S2251171719500041},
  url = {https://doi.org/10.48550/arXiv.1806.09531},
  eprint = {arXiv:1806.09531},
}

@article{voytek2014probing,
  title={Probing the dark ages at z~ 20: The SCI-HI 21 cm all-sky spectrum experiment},
  author={Voytek, Tabitha C and Natarajan, Aravind and Garc{\'\i}a, Jos{\'e} Miguel J{\'a}uregui and Peterson, Jeffrey B and L{\'o}pez-Cruz, Omar},
  journal={The Astrophysical Journal Letters},
  volume={782},
  number={1},
  pages={L9},
  year={2014},
  publisher={IOP Publishing},
  doi={10.1088/2041-8205/782/1/L9},
  url = {https://doi.org/10.48550/arXiv.1311.0014},
  eprint = {arXiv:1311.0014},
}

@article{sokolowski2015bighorns,
  title={BIGHORNS-broadband instrument for global HydrOgen ReioNisation signal},
  author={Sokolowski, Marcin and Tremblay, Steven E and Wayth, Randall B and Tingay, Steven J and Clarke, Nathan and Roberts, Paul and Waterson, Mark and Ekers, Ronald D and Hall, Peter and Lewis, Morgan and others},
  journal={Publications of the Astronomical Society of Australia},
  volume={32},
  pages={e004},
  year={2015},
  publisher={Cambridge University Press},
  doi={https://doi.org/10.1017/pasa.2015.3},
  url = {https://doi.org/10.48550/arXiv.1501.02922},
  eprint = {arXiv:1501.02922},
}

@article{price2018design,
    author = {Price, D C and Greenhill, L J and Fialkov, A and Bernardi, G and Garsden, H and Barsdell, B R and Kocz, J and Anderson, M M and Bourke, S A and Craig, J and Dexter, M R and Dowell, J and Eastwood, M W and Eftekhari, T and Ellingson, S W and Hallinan, G and Hartman, J M and Kimberk, R and Lazio, T Joseph W and Leiker, S and MacMahon, D and Monroe, R and Schinzel, F and Taylor, G B and Tong, E and Werthimer, D and Woody, D P},
    title = {Design and characterization of the Large-aperture Experiment to Detect the Dark Age (LEDA) radiometer systems},
    journal = {Monthly Notices of the Royal Astronomical Society},
    volume = {478},
    number = {3},
    pages = {4193-4213},
    year = {2018},
    month = {05},
    issn = {0035-8711},
    doi = {10.1093/mnras/sty1244},
    url = {https://doi.org/10.48550/arXiv.1709.09313},
    eprint = {arXiv:1709.09313},
}

@article{mckinley2020all,
    author = {McKinley, B and Trott, C M and Sokolowski, M and Wayth, R B and Sutinjo, A and Patra, N and NambissanT., J and Ung, D C X},
    title = {The All-Sky SignAl Short-Spacing INterferometer (ASSASSIN) – I. Global-sky measurements with the Engineering Development Array-2},
    journal = {Monthly Notices of the Royal Astronomical Society},
    volume = {499},
    number = {1},
    pages = {52-67},
    year = {2020},
    month = {09},
    issn = {0035-8711},
    doi = {10.1093/mnras/staa2804},
    url = {https://doi.org/10.48550/arXiv.2009.06146},
    eprint = {arXiv:2009.06146},
}

@article{burns2021global,
  title={Global 21-cm cosmology from the farside of the Moon},
  author={Burns, Jack and Bale, Stuart and Bradley, Richard and Ahmed, Z and Allen, SW and Bowman, J and Furlanetto, S and MacDowall, R and Mirocha, J and Nhan, B and others},
  year={2021},
  url={https://arxiv.org/abs/2103.05085},
  eprint = {arXiv:2103.05085},
}

@article{chen2021discovering,
  title={Discovering the sky at the longest wavelengths with a lunar orbit array},
  author={Chen, Xuelei and Yan, Jingye and Deng, Li and Wu, Fengquan and Wu, Lin and Xu, Yidong and Zhou, Li},
  journal={Philosophical Transactions of the Royal Society A},
  volume={379},
  number={2188},
  pages={20190566},
  year={2021},
  publisher={The Royal Society Publishing},
  doi={https://doi.org/10.1098/rsta.2019.0566},
  url={https://doi.org/10.48550/arXiv.2007.15794},
  eprint = {arXiv:2007.15794},
}

@article{de2019reach,
  title={The REACH radiometer for detecting the 21-cm hydrogen signal from redshift $z \approx 7.5$--28},
  author={de Lera Acedo, E and de Villiers, DIL and Razavi-Ghods, N and Handley, W and Fialkov, A and Magro, Alessio and Anstey, D and Bevins, HTJ and Chiello, R and Cumner, J and others},
  journal={Nature Astronomy},
  volume={6},
  number={8},
  pages={984--998},
  year={2022},
  publisher={Nature Publishing Group UK London},
  doi={https://doi.org/10.1038/s41550-022-01709-9},
  url={https://doi.org/10.48550/arXiv.2210.07409},
  eprint = {arXiv:2210.07409},
}

@article{deboer2017hydrogen,
  title={Hydrogen epoch of reionization array (HERA)},
  author={DeBoer, David R and Parsons, Aaron R and Aguirre, James E and Alexander, Paul and Ali, Zaki S and Beardsley, Adam P and Bernardi, Gianni and Bowman, Judd D and Bradley, Richard F and Carilli, Chris L and others},
  journal={Publications of the Astronomical Society of the Pacific},
  volume={129},
  number={974},
  pages={045001},
  year={2017},
  publisher={IOP Publishing},
  doi={https://doi.org/10.1088/1538-3873/129/974/045001},
  url={https://doi.org/10.48550/arXiv.1606.07473},
  eprint = {arXiv:1606.07473},
}

@article{singh2019redshifted,
  title={The redshifted 21 cm signal in the EDGES low-band spectrum},
  author={Singh, Saurabh and Subrahmanyan, Ravi},
  journal={The Astrophysical Journal},
  volume={880},
  number={1},
  pages={26},
  year={2019},
  publisher={IOP Publishing},
  doi={https://doi.org/10.3847/1538-4357/ab2879},
  url={https://doi.org/10.48550/arXiv.1903.04540},
  eprint = {arXiv:1903.04540}
}

@article{driskell2022structure,
  title={Structure formation and the global 21-cm signal in the presence of Coulomb-like dark matter-baryon interactions},
  author={Driskell, Trey and Nadler, Ethan O and Mirocha, Jordan and Benson, Andrew and Boddy, Kimberly K and Morton, Timothy D and Lashner, Jack and An, Rui and Gluscevic, Vera},
  journal={Physical Review D},
  volume={106},
  number={10},
  pages={103525},
  year={2022},
  publisher={APS},
  doi={https://doi.org/10.1103/PhysRevD.106.103525},
  url={https://doi.org/10.48550/arXiv.2209.04499},
  eprint = {arXiv:2209.04499}
}

@article{aghanim2020planck,
  title={Planck 2018 results-VI. Cosmological parameters},
  author={Aghanim, Nabila and Akrami, Yashar and Ashdown, Mark and Aumont, J and Baccigalupi, C and Ballardini, M and Banday, AJ and Barreiro, RB and Bartolo, N and Basak, S and others},
  journal={Astronomy \& Astrophysics},
  volume={641},
  pages={A6},
  year={2020},
  publisher={EDP sciences},
  doi={https://doi.org/10.1051/0004-6361/201833910},
  url={https://doi.org/10.48550/arXiv.1807.06209},
  eprint = {arXiv:1807.06209}
}

@article{hibbard2022constraining,
  title={Constraining Warm Dark Matter and Population III Stars with the Global 21 cm Signal},
  author={Hibbard, Joshua J and Mirocha, Jordan and Rapetti, David and Bassett, Neil and Burns, Jack O and Tauscher, Keith},
  journal={The Astrophysical Journal},
  volume={929},
  number={2},
  pages={151},
  year={2022},
  publisher={IOP Publishing},
  doi={https://doi.org/10.3847/1538-4357/ac5ea3},
  url={https://doi.org/10.48550/arXiv.2201.02638},
  eprint = {arXiv:2201.02638}
}

@article{haslam1982408,
  doi = {https://doi.org/10.1051/0004-6361/200912161},
  title={A 408 MHz all-sky continuum survey. II-The atlas of contour maps},
  author={Haslam, CGT and Salter, CJ and Stoffel, H and Wilson, WEz},
  journal={Astronomy and Astrophysics Supplement Series, vol. 47, Jan. 1982, p. 1, 2, 4-51, 53-142.},
  volume={47},
  pages={1},
  year={1982}
}

@article{dvorkin2014constraining,
  title={Constraining dark matter-baryon scattering with linear cosmology},
  author={Dvorkin, Cora and Blum, Kfir and Kamionkowski, Marc},
  journal={Physical Review D},
  volume={89},
  number={2},
  pages={023519},
  year={2014},
  publisher={APS},
  doi={https://doi.org/10.1103/PhysRevD.89.023519},
  url={https://doi.org/10.48550/arXiv.1311.2937},
  eprint = {arXiv:1311.2937}
}

@article{slatyer2018early,
  title={Early-Universe constraints on dark matter-baryon scattering and their implications for a global 21 cm signal},
  author={Slatyer, Tracy R and Wu, Chih-Liang},
  journal={Physical Review D},
  volume={98},
  number={2},
  pages={023013},
  year={2018},
  publisher={APS},
  doi={https://doi.org/10.1103/PhysRevD.98.023013},
  url={https://doi.org/10.48550/arXiv.1803.09734},
  eprint = {arXiv:1803.09734}
}

@article{gluscevic2018constraints,
  title={Constraints on scattering of keV--TeV dark matter with protons in the early Universe},
  author={Gluscevic, Vera and Boddy, Kimberly K},
  journal={Physical Review Letters},
  volume={121},
  number={8},
  pages={081301},
  year={2018},
  publisher={APS},
  doi={https://doi.org/10.1103/PhysRevLett.121.081301},
  url={https://doi.org/10.48550/arXiv.1712.07133},
  eprint = {arXiv:1712.07133}
}

@article{boddy2018first,
  title={First cosmological constraint on the effective theory of dark matter-proton interactions},
  author={Boddy, Kimberly K and Gluscevic, Vera},
  journal={Physical Review D},
  volume={98},
  number={8},
  pages={083510},
  year={2018},
  publisher={APS},
  doi={https://doi.org/10.1103/PhysRevD.98.083510},
  url={https://doi.org/10.48550/arXiv.1801.08609},
  eprint = {arXiv:1801.08609}
}

@article{boddy2018critical,
  title={Critical assessment of CMB limits on dark matter-baryon scattering: New treatment of the relative bulk velocity},
  author={Boddy, Kimberly K and Gluscevic, Vera and Poulin, Vivian and Kovetz, Ely D and Kamionkowski, Marc and Barkana, Rennan},
  journal={Physical Review D},
  volume={98},
  number={12},
  pages={123506},
  year={2018},
  publisher={APS},
  doi={https://doi.org/10.1103/PhysRevD.98.123506},
  url={https://doi.org/10.48550/arXiv.1808.00001},
  eprint = {arXiv:1808.00001}
}

@article{nguyen2021observational,
  title={Observational constraints on dark matter scattering with electrons},
  author={Nguyen, David V and Sarnaaik, Dimple and Boddy, Kimberly K and Nadler, Ethan O and Gluscevic, Vera},
  journal={Physical Review D},
  volume={104},
  number={10},
  pages={103521},
  year={2021},
  publisher={APS},
  doi={https://doi.org/10.1103/PhysRevD.104.103521},
  url={https://doi.org/10.48550/arXiv.2107.12380},
  eprint = {arXiv:2107.12380}
}

@article{maamari2021bounds,
  title={Bounds on velocity-dependent dark matter--proton scattering from Milky Way satellite abundance},
  author={Maamari, Karime and Gluscevic, Vera and Boddy, Kimberly K and Nadler, Ethan O and Wechsler, Risa H},
  journal={The Astrophysical Journal Letters},
  volume={907},
  number={2},
  pages={L46},
  year={2021},
  publisher={IOP Publishing},
  doi={https://doi.org/10.3847/2041-8213/abd807},
  url={https://doi.org/10.48550/arXiv.2010.02936},
  eprint = {arXiv:2010.02936}
}

@article{rogers2022limits,
  title={Limits on the Light Dark Matter--Proton Cross Section from Cosmic Large-Scale Structure},
  author={Rogers, Keir K and Dvorkin, Cora and Peiris, Hiranya V},
  journal={Physical Review Letters},
  volume={128},
  number={17},
  pages={171301},
  year={2022},
  publisher={APS},
  doi={https://doi.org/10.1103/PhysRevLett.128.171301},
  url={https://doi.org/10.48550/arXiv.2111.10386},
  eprint = {arXiv:2111.10386}
}

@article{becker2021cosmological,
  title={Cosmological constraints on multi-interacting dark matter},
  author={Becker, Niklas and Hooper, Deanna C and Kahlhoefer, Felix and Lesgourgues, Julien and Sch{\"o}neberg, Nils},
  journal={Journal of Cosmology and Astroparticle Physics},
  volume={2021},
  number={02},
  pages={019},
  year={2021},
  publisher={IOP Publishing},
  doi={https://doi.org/10.1088/1475-7516/2021/02/019},
  url={https://doi.org/10.48550/arXiv.2010.04074},
  eprint = {arXiv:2010.04074}
}

@article{nadler2019constraints,
  title={Constraints on dark matter microphysics from the milky way satellite population},
  author={Nadler, Ethan O and Gluscevic, Vera and Boddy, Kimberly K and Wechsler, Risa H},
  journal={The Astrophysical Journal Letters},
  volume={878},
  number={2},
  pages={L32},
  year={2019},
  publisher={IOP Publishing},
  doi={https://doi.org/10.3847/2041-8213/ab1eb2},
  url={https://doi.org/10.48550/arXiv.1904.10000},
  eprint = {arXiv:1904.10000}
}

@article{nadler2021constraints,
  title={Constraints on dark matter properties from observations of Milky Way satellite galaxies},
  author={Nadler, EO and Drlica-Wagner, A and Bechtol, K and Mau, S and Wechsler, RH and Gluscevic, V and Boddy, K and Pace, AB and Li, TS and McNanna, M and others},
  journal={Physical review letters},
  volume={126},
  number={9},
  pages={091101},
  year={2021},
  publisher={APS},
  doi={https://doi.org/10.1103/PhysRevLett.126.091101},
  url={https://doi.org/10.48550/arXiv.2008.00022},
  eprint = {arXiv:2008.00022}
}

@article{li2023atacama,
  title={The Atacama Cosmology Telescope: limits on dark matter-baryon interactions from DR4 power spectra},
  author={Li, Zack and An, Rui and Gluscevic, Vera and Boddy, Kimberly K and Bond, J Richard and Calabrese, Erminia and Dunkley, Jo and Gallardo, Patricio A and Guan, Yilun and Hincks, Adam and others},
  journal={Journal of Cosmology and Astroparticle Physics},
  volume={2023},
  number={02},
  pages={046},
  year={2023},
  publisher={IOP Publishing},
  doi={https://doi.org/10.1088/1475-7516/2023/02/046},
  url = {https://arxiv.org/abs/2208.08985},
  eprint = {arXiv:2208.08985}
}

@article{boddy2022investigation,
  title={Investigation of CMB constraints for dark matter-helium scattering},
  author={Boddy, Kimberly K and Krnjaic, Gordan and Moltner, Stacie},
  journal={Physical Review D},
  volume={106},
  number={4},
  pages={043510},
  year={2022},
  publisher={APS},
  doi={https://doi.org/10.1103/PhysRevD.106.043510},
  url = {https://arxiv.org/abs/2204.04225},
  eprint = {arXiv:2204.04225}
}

@article{lin2023dark,
  title={Dark Matter Subhalo Evaporation by Coulomb-like Interaction with Galactic Gas},
  author={Lin, Yugen and Gao, Yu},
  year={2023},
  url = {https://doi.org/10.48550/arXiv.2311.11584},
  eprint = {arXiv:2311.11584},
}

@article{press1974formation,
  url = {https://ui.adsabs.harvard.edu/abs/1974ApJ...187..425P/abstract},
  title={Formation of galaxies and clusters of galaxies by self-similar gravitational condensation},
  author={Press, William H and Schechter, Paul},
  journal={Astrophysical Journal, Vol. 187, pp. 425-438 (1974)},
  volume={187},
  pages={425--438},
  year={1974},
  doi = {10.1086/152650},
}

@article{bond1991excursion,
  url = {https://ui.adsabs.harvard.edu/abs/1991ApJ...379..440B/abstract},
  title={Excursion set mass functions for hierarchical Gaussian fluctuations},
  author={Bond, JR and Cole, Shaun and Efstathiou, George and Kaiser, Nick},
  journal={Astrophysical Journal},
  volume={379},
  pages={440--460},
  year={1991},
  doi = {10.1086/170520},
}

@article{tinker2008toward,
  title={Toward a halo mass function for precision cosmology: The Limits of universality},
  author={Tinker, Jeremy and Kravtsov, Andrey V and Klypin, Anatoly and Abazajian, Kevork and Warren, Michael and Yepes, Gustavo and Gottl{\"o}ber, Stefan and Holz, Daniel E},
  journal={The Astrophysical Journal},
  volume={688},
  number={2},
  pages={709},
  year={2008},
  publisher={IOP Publishing},
  doi={https://doi.org/10.1086/591439},
  url = {https://doi.org/10.48550/arXiv.0803.2706},
  eprint = {arXiv:0803.2706},
}

@article{despali2016universality,
  title={The universality of the virial halo mass function and models for non-universality of other halo definitions},
  author={Despali, Giulia and Giocoli, Carlo and Angulo, Raul E and Tormen, Giuseppe and Sheth, Ravi K and Baso, Giacomo and Moscardini, Lauro},
  journal={Monthly Notices of the Royal Astronomical Society},
  volume={456},
  number={3},
  pages={2486--2504},
  year={2016},
  publisher={The Royal Astronomical Society},
  doi={https://doi.org/10.1093/mnras/stv2842},
  url = {https://doi.org/10.48550/arXiv.1507.05627},
  eprint = {arXiv:1507.05627},
}

@article{ondaro2022non,
  title={Non-universality of the mass function: dependence on the growth rate and power spectrum shape},
  author={Ondaro-Mallea, Lurdes and Angulo, Raul E and Zennaro, Matteo and Contreras, Sergio and Aric{\`o}, Giovanni},
  journal={Monthly Notices of the Royal Astronomical Society},
  volume={509},
  number={4},
  pages={6077--6090},
  year={2022},
  publisher={Oxford University Press},
  doi={https://doi.org/10.1093/mnras/stab3337},
  url = {https://doi.org/10.48550/arXiv.2102.08958},
  eprint = {arXiv:2102.08958},
}

@article{furlanetto2006global,
  title={The global 21-centimeter background from high redshifts},
  author={Furlanetto, Steven R},
  journal={Monthly Notices of the Royal Astronomical Society},
  volume={371},
  number={2},
  pages={867--878},
  year={2006},
  publisher={Blackwell Publishing Ltd Oxford, UK},
  doi={https://doi.org/10.1111/j.1365-2966.2006.10725.x},
  url = {https://arxiv.org/abs/astro-ph/0604040},
  eprint = {arXiv:astro-ph/0604040},
}

@article{schneider2015structure,
  title={Structure formation with suppressed small-scale perturbations},
  author={Schneider, Aurel},
  journal={Monthly Notices of the Royal Astronomical Society},
  volume={451},
  number={3},
  pages={3117--3130},
  year={2015},
  publisher={Oxford University Press},
  doi = {10.1093/mnras/stv1169},
  url = {https://doi.org/10.48550/arXiv.1412.2133},
  eprint = {arXiv:1412.2133},
}

@article{mineo2012x,
  title={X-ray emission from star-forming galaxies--I. High-mass X-ray binaries},
  author={Mineo, S and Gilfanov, M and Sunyaev, R},
  journal={Monthly Notices of the Royal Astronomical Society},
  volume={419},
  number={3},
  pages={2095--2115},
  year={2012},
  publisher={The Royal Astronomical Society},
  doi = {https://doi.org/10.1111/j.1365-2966.2011.19862.x},
  url = {https://doi.org/10.48550/arXiv.1105.4610},
  eprint = {arXiv:1105.4610},
}

@article{zwicky2009republication,
  doi = {https://doi.org/10.1007/s10714-008-0707-4},
  title={Republication of: The redshift of extragalactic nebulae},
  author={Zwicky, Fritz},
  journal={General Relativity and Gravitation},
  volume={41},
  number={1},
  pages={207--224},
  year={2009},
  publisher={Springer},
}

@article{rubin1970rotation,
  url = {https://ui.adsabs.harvard.edu/abs/1970ApJ...159..379R/abstract},
  title={Rotation of the Andromeda nebula from a spectroscopic survey of emission regions},
  author={Rubin, Vera C and Ford Jr, W Kent},
  journal={Astrophysical Journal},
  volume={159},
  pages={379},
  year={1970},
  doi = {10.1086/150317},
}

@article{refregier2003weak,
  title={Weak gravitational lensing by large-scale structure},
  author={Refregier, Alexandre},
  journal={Annual Review of Astronomy and Astrophysics},
  volume={41},
  number={1},
  pages={645--668},
  year={2003},
  publisher={Annual Reviews 4139 El Camino Way, PO Box 10139, Palo Alto, CA 94303-0139, USA},
  doi = {https://doi.org/10.1146/annurev.astro.41.111302.102207},
  url = {https://doi.org/10.48550/arXiv.astro-ph/0307212},
  eprint = {arXiv:astro-ph/0307212},
}

@article{clowe2006direct,
  title={A direct empirical proof of the existence of dark matter},
  author={Clowe, Douglas and Brada{\v{c}}, Maru{\v{s}}a and Gonzalez, Anthony H and Markevitch, Maxim and Randall, Scott W and Jones, Christine and Zaritsky, Dennis},
  journal={The Astrophysical Journal},
  volume={648},
  number={2},
  pages={L109},
  year={2006},
  publisher={IOP Publishing},
  doi= {https://doi.org/10.1086/508162},
  url = {https://doi.org/10.48550/arXiv.astro-ph/0608407},
  eprint = {arXiv:astro-ph/0608407},
}

@article{abdalla2022cosmology,
  title={Cosmology intertwined: A review of the particle physics, astrophysics, and cosmology associated with the cosmological tensions and anomalies},
  author={Abdalla, Elcio and Abell{\'a}n, Guillermo Franco and Aboubrahim, Amin and Agnello, Adriano and Akarsu, {\"O}zg{\"u}r and Akrami, Yashar and Alestas, George and Aloni, Daniel and Amendola, Luca and Anchordoqui, Luis A and others},
  journal={Journal of High Energy Astrophysics},
  volume={34},
  pages={49--211},
  year={2022},
  publisher={Elsevier},
  doi={https://doi.org/10.1016/j.jheap.2022.04.002},
  url = {https://doi.org/10.48550/arXiv.2203.06142},
  eprint = {arXiv:2203.06142},
}

@article{anchordoqui2021dissecting,
  title={Dissecting the $\mathrm{H}_0$ and $\mathrm{S}_8$ tensions with Planck+ BAO+ supernova type Ia in multi-parameter cosmologies},
  author={Anchordoqui, Luis A and Di Valentino, Eleonora and Pan, Supriya and Yang, Weiqiang},
  journal={Journal of High Energy Astrophysics},
  volume={32},
  pages={28--64},
  year={2021},
  publisher={Elsevier}, 
  doi={https://doi.org/10.1016/j.jheap.2021.08.001},
  url = {https://doi.org/10.48550/arXiv.2107.13932},
  eprint = {arXiv:2107.13932},
}

@article{di2021realm,
  title={In the realm of the Hubble tension—a review of solutions},
  author={Di Valentino, Eleonora and Mena, Olga and Pan, Supriya and Visinelli, Luca and Yang, Weiqiang and Melchiorri, Alessandro and Mota, David F and Riess, Adam G and Silk, Joseph},
  journal={Classical and Quantum Gravity},
  volume={38},
  number={15},
  pages={153001},
  year={2021},
  publisher={IOP Publishing},
  doi={https://doi.org/10.1088/1361-6382/ac086d},
  url = {https://doi.org/10.48550/arXiv.2103.01183},
  eprint = {arXiv:2103.01183},
}

@article{schoneberg2022h0,
  title={The H0 Olympics: A fair ranking of proposed models},
  author={Sch{\"o}neberg, Nils and Abell{\'a}n, Guillermo Franco and S{\'a}nchez, Andrea P{\'e}rez and Witte, Samuel J and Poulin, Vivian and Lesgourgues, Julien},
  journal={Physics Reports},
  volume={984},
  pages={1--55},
  year={2022},
  publisher={Elsevier},
  doi={https://doi.org/10.1016/j.physrep.2022.07.001},
  url = {https://doi.org/10.48550/arXiv.2107.10291},
  eprint = {arXiv:2107.10291},
}

@article{perivolaropoulos2022challenges,
  title={Challenges for {$\Lambda$CDM}: An update},
  author={Perivolaropoulos, Leandros and Skara, Foteini},
  journal={New Astronomy Reviews},
  volume={95},
  pages={101659},
  year={2022},
  publisher={Elsevier},
  doi={https://doi.org/10.1016/j.newar.2022.101659},
  url = {https://doi.org/10.48550/arXiv.2105.05208},
  eprint = {arXiv:2105.05208},
}

@article{escudero2018fresh,
  title={A fresh look into the interacting dark matter scenario},
  author={Escudero, Miguel and Lopez-Honorez, Laura and Mena, Olga and Palomares-Ruiz, Sergio and Villanueva-Domingo, Pablo},
  journal={Journal of Cosmology and Astroparticle Physics},
  volume={2018},
  number={06},
  pages={007},
  year={2018},
  publisher={IOP Publishing},
  doi={https://doi.org/10.1088/1475-7516/2018/06/007},
  url = {https://doi.org/10.48550/arXiv.1803.08427},
  eprint = {arXiv:1803.08427},
}

@article{burns2019dark,
  url = {https://doi.org/10.48550/arXiv.1902.06147},
  eprint = {arXiv:1902.06147},
  title={Dark cosmology: investigating dark matter \& exotic physics in the dark ages using the redshifted 21-cm global spectrum},
  author={Burns, Jack O and Bale, Stuart and Bassett, Neil and Bowman, Judd and Bradley, Richard and Fialkov, Anastasia and Furlanetto, Steven and Hecht, Michael and Klein-Wolt, Marc and Lonsdale, Colin and others},
  year={2019}
}

@article{adams2023improved,
  title={Improved Constraints on the 21 cm EoR Power Spectrum and the X-Ray Heating of the IGM with HERA Phase I Observations},
  author={Adams, Tyrone and Aguirre, James E and Alexander, Paul and Ali, Zaki S and Baartman, Rushelle and Balfour, Yanga and Barkana, Rennan and Beardsley, Adam P and Bernardi, Gianni and Billings, Tashalee S and others},
  journal={The Astrophysical Journal},
  volume={945},
  number={2},
  pages={124},
  year={2023},
  publisher={IOP Publishing},
  doi={https://doi.org/10.3847/1538-4357/acaf50},
  url = {https://doi.org/10.48550/arXiv.2210.04912},
  eprint = {arXiv:2210.04912},
}

@article{sims2019joint,
  title={Joint estimation of the Epoch of Reionization power spectrum and foregrounds},
  author={Sims, Peter H and Pober, Jonathan C},
  journal={Monthly Notices of the Royal Astronomical Society},
  volume={488},
  number={2},
  pages={2904--2916},
  year={2019},
  publisher={Oxford University Press},
  doi={https://doi.org/10.1093/mnras/stz1888},
  url = {https://doi.org/10.48550/arXiv.1907.02608},
  eprint = {arXiv:1907.02608},
}

@article{bradley2019ground,
  title={A ground plane artifact that induces an absorption profile in averaged spectra from global 21 cm measurements, with possible application to EDGES},
  author={Bradley, Richard F and Tauscher, Keith and Rapetti, David and Burns, Jack O},
  journal={The Astrophysical Journal},
  volume={874},
  number={2},
  pages={153},
  year={2019},
  publisher={IOP Publishing},
  doi={https://doi.org/10.3847/1538-4357/ab0d8b},
  url = {https://doi.org/10.48550/arXiv.1810.09015},
  eprint = {arXiv:1810.09015},
}

@article{tauscher2020global,
  title={Global 21 cm Signal Extraction from Foreground and Instrumental Effects. III. Utilizing Drift-scan Time Dependence and Full Stokes Measurements},
  author={Tauscher, Keith and Rapetti, David and Burns, Jack O},
  journal={The Astrophysical Journal},
  volume={897},
  number={2},
  pages={175},
  year={2020},
  publisher={IOP Publishing},
  doi = {10.3847/1538-4357/ab9b2a},
  url = {https://doi.org/10.48550/arXiv.2003.05452},
  eprint = {arXiv:2003.05452},
}

@article{benson2013dark,
  title={Dark matter halo merger histories beyond cold dark matter--I. Methods and application to warm dark matter},
  author={Benson, Andrew J and Farahi, Arya and Cole, Shaun and Moustakas, Leonidas A and Jenkins, Adrian and Lovell, Mark and Kennedy, Rachel and Helly, John and Frenk, Carlos},
  journal={Monthly Notices of the Royal Astronomical Society},
  volume={428},
  number={2},
  pages={1774--1789},
  year={2013},
  publisher={Oxford University Press},
  doi = {https://doi.org/10.1093/mnras/sts159},
  url = {https://doi.org/10.48550/arXiv.1209.3018},
  eprint = {arXiv:1209.3018},
}

@article{barkana2018possible,
  title={Possible interaction between baryons and dark-matter particles revealed by the first stars},
  author={Barkana, Rennan},
  journal={Nature},
  volume={555},
  number={7694},
  pages={71--74},
  year={2018},
  publisher={Nature Publishing Group UK London},
  doi = {https://doi.org/10.1038/nature25791},
}

@article{berlin2018severely,
  title={Severely constraining dark-matter interpretations of the 21-cm anomaly},
  author={Berlin, Asher and Hooper, Dan and Krnjaic, Gordan and McDermott, Samuel D},
  journal={Physical review letters},
  volume={121},
  number={1},
  pages={011102},
  year={2018},
  publisher={APS},
  doi = {https://doi.org/10.1103/PhysRevLett.121.011102},
  url = {https://doi.org/10.48550/arXiv.1803.02804},
  eprint = {arXiv:1803.02804},
}

@article{fialkov2018constraining,
  title={Constraining baryon--dark-matter scattering with the cosmic dawn 21-cm signal},
  author={Fialkov, Anastasia and Barkana, Rennan and Cohen, Aviad},
  journal={Physical review letters},
  volume={121},
  number={1},
  pages={011101},
  year={2018},
  publisher={APS},
  doi = {https://doi.org/10.1103/PhysRevLett.121.011101},
  url = {https://doi.org/10.48550/arXiv.1802.10577},
  eprint = {arXiv:1802.10577},
}

@article{tulin2018dark,
  title={Dark matter self-interactions and small scale structure},
  author={Tulin, Sean and Yu, Hai-Bo},
  journal={Physics Reports},
  volume={730},
  pages={1--57},
  year={2018},
  publisher={Elsevier},
  doi = {https://doi.org/10.1016/j.physrep.2017.11.004},
  url = {https://doi.org/10.48550/arXiv.1705.02358},
  eprint = {arXiv:1705.02358},
}

@article{zentner2022critical,
  title={A critical assessment of solutions to the galaxy diversity problem},
  author={Zentner, Aidan and Dandavate, Siddharth and Slone, Oren and Lisanti, Mariangela},
  journal={Journal of Cosmology and Astroparticle Physics},
  volume={2022},
  number={07},
  pages={031},
  year={2022},
  publisher={IOP Publishing},
  doi = {https://doi.org/10.1088/1475-7516/2022/07/031},
  url = {https://doi.org/10.48550/arXiv.2202.00012},
  eprint = {arXiv:2202.00012},
}

@article{bowman2008toward,
  title={Toward empirical constraints on the global redshifted 21 cm brightness temperature during the epoch of reionization},
  author={Bowman, Judd D and Rogers, Alan EE and Hewitt, Jacqueline N},
  journal={The Astrophysical Journal},
  volume={676},
  number={1},
  pages={1},
  year={2008},
  publisher={IOP Publishing},
  doi = {https://doi.org/10.1086/528675},
  url = {https://doi.org/10.48550/arXiv.0710.2541},
  eprint = {arXiv:0710.2541},
}

@article{bowman2010lower,
  doi = {https://doi.org/10.1038/nature09601},
  title={A lower limit of $\Delta$ z> 0.06 for the duration of the reionization epoch},
  author={Bowman, Judd D and Rogers, Alan EE},
  journal={Nature},
  volume={468},
  number={7325},
  pages={796--798},
  year={2010},
  publisher={Nature Publishing Group UK London}
}

@article{fisher1935logic,
  doi = {https://doi.org/10.2307/2342435},
  title={The logic of inductive inference},
  author={Fisher, Ronald A},
  journal={Journal of the royal statistical society},
  volume={98},
  number={1},
  pages={39--82},
  year={1935},
  publisher={JSTOR}
}

@article{ade2019simons,
  title={The Simons Observatory: science goals and forecasts},
  author={Ade, Peter and Aguirre, James and Ahmed, Zeeshan and Aiola, Simone and Ali, Aamir and Alonso, David and Alvarez, Marcelo A and Arnold, Kam and Ashton, Peter and Austermann, Jason and others},
  journal={Journal of Cosmology and Astroparticle Physics},
  volume={2019},
  number={02},
  pages={056},
  year={2019},
  publisher={IOP Publishing},
  doi = {https://doi.org/10.1088/1475-7516/2019/02/056},
  url = {https://doi.org/10.48550/arXiv.1808.07445},
  eprint = {arXiv:1808.07445},
}

@article{coe2009fisher,
  url = {https://doi.org/10.48550/arXiv.0906.4123},
  eprint = {arXiv:0906.4123},
  title={Fisher matrices and confidence ellipses: a quick-start guide and software},
  author={Coe, Dan},
  year={2009}
}

@article{munoz2015heating,
  title={Heating of baryons due to scattering with dark matter during the dark ages},
  author={Mu{\~n}oz, Julian B and Kovetz, Ely D and Ali-Ha{\"\i}moud, Yacine},
  journal={Physical Review D},
  volume={92},
  number={8},
  pages={083528},
  year={2015},
  publisher={APS},
  doi = {https://doi.org/10.1103/PhysRevD.92.083528},
  url = {https://doi.org/10.48550/arXiv.1509.00029},
  eprint = {arXiv:1509.00029},
}

@article{ali2015constraints,
  title={Constraints on dark matter interactions with standard model particles from cosmic microwave background spectral distortions},
  author={Ali-Ha{\"\i}moud, Yacine and Chluba, Jens and Kamionkowski, Marc},
  journal={Physical Review Letters},
  volume={115},
  number={7},
  pages={071304},
  year={2015},
  publisher={APS},
  doi = {https://doi.org/10.1103/PhysRevLett.115.071304},
  url = {https://doi.org/10.48550/arXiv.1506.04745},
  eprint = {arXiv:1506.04745},
}

@article{kovetz2018tighter,
  title={Tighter limits on dark matter explanations of the anomalous EDGES 21 cm signal},
  author={Kovetz, Ely D and Poulin, Vivian and Gluscevic, Vera and Boddy, Kimberly K and Barkana, Rennan and Kamionkowski, Marc},
  journal={Physical Review D},
  volume={98},
  number={10},
  pages={103529},
  year={2018},
  publisher={APS},
  doi = {https://doi.org/10.1103/PhysRevD.98.103529},
  url = {https://doi.org/10.48550/arXiv.1807.11482},
  eprint = {arXiv:1807.11482},
}

@article{mondal2024constraining,
  title={Constraining exotic dark matter models with the dark ages 21-cm signal},
  author={Mondal, Rajesh and Barkana, Rennan and Fialkov, Anastasia},
  journal={Monthly Notices of the Royal Astronomical Society},
  volume={527},
  number={1},
  pages={1461--1471},
  year={2024},
  publisher={Oxford University Press},
  doi = {https://doi.org/10.1093/mnras/stad3317},
  url = {https://doi.org/10.48550/arXiv.2310.15530},
  eprint = {arXiv:2310.15530},
}

@article{das2022modified,
  title={Modified dispersion relations and a potential explanation of the EDGES anomaly},
  author={Das, Saurya and Fridman, Mitja and Lambiase, Gaetano and Stabile, Antonio and Vagenas, Elias C},
  journal={The European Physical Journal C},
  volume={82},
  number={8},
  pages={1--10},
  year={2022},
  publisher={Springer},
  doi = {https://doi.org/10.1140/epjc/s10052-022-10680-8},
  url = {https://doi.org/10.48550/arXiv.2110.02340},
  eprint = {arXiv:2110.02340},
}

@article{gluscevic2019cosmological,
  url = {https://arxiv.org/abs/1903.05140},
  eprint = {arXiv:1903.05140},
  title={Cosmological probes of dark matter interactions: The next decade},
  author={Gluscevic, Vera and Ali-Haimoud, Yacine and Bechtol, Keith and Boddy, Kimberly K and B{\oe}hm, C{\'e}line and Chluba, Jens and Cyr-Racine, Francis-Yan and Dvorkin, Cora and Grin, Daniel and Lesgourgues, Julien and others},
  year={2019}
}

@article{slatyer2009cmb,
  title={CMB constraints on WIMP annihilation: energy absorption during the recombination epoch},
  author={Slatyer, Tracy R and Padmanabhan, Nikhil and Finkbeiner, Douglas P},
  journal={Physical Review D},
  volume={80},
  number={4},
  pages={043526},
  year={2009},
  publisher={APS},
  doi = {https://doi.org/10.1103/PhysRevD.80.043526},
  url = {https://doi.org/10.48550/arXiv.0906.1197},
  eprint = {arXiv:0906.1197},
}

@article{xu2018probing,
  title={Probing sub-GeV dark matter-baryon scattering with cosmological observables},
  author={Xu, Weishuang Linda and Dvorkin, Cora and Chael, Andrew},
  journal={Physical Review D},
  volume={97},
  number={10},
  pages={103530},
  year={2018},
  publisher={APS},
  doi = {https://doi.org/10.1103/PhysRevD.97.103530},
  url = {https://doi.org/10.48550/arXiv.1802.06788},
  eprint = {arXiv:1802.06788},
}

@article{ooba2019cosmological,
  title={Cosmological constraints on the velocity-dependent baryon-dark matter coupling},
  author={Ooba, Junpei and Tashiro, Hiroyuki and Kadota, Kenji},
  journal={Journal of Cosmology and Astroparticle Physics},
  volume={2019},
  number={09},
  pages={020},
  year={2019},
  publisher={IOP Publishing},
  doi = {https://doi.org/10.1088/1475-7516/2019/09/020},
  url = {https://doi.org/10.48550/arXiv.1902.00826},
  eprint = {arXiv:1902.00826},
}

@book{rohlfs2013tools,
    author = {{Wilson}, Thomas L. and {Rohlfs}, Kristen and {H{\"u}ttemeister}, Susanne},
    title = "{Tools of Radio Astronomy}",
    year = 2013,
    publisher = "Springer Berlin, Heidelberg",
    doi = {10.1007/978-3-642-39950-3},
    adsurl = {https://ui.adsabs.harvard.edu/abs/2013tra..book.....W}
}

@article{furlanetto2006cosmology,
  title={Cosmology at low frequencies: The 21 cm transition and the high-redshift Universe},
  author={Furlanetto, Steven R and Oh, S Peng and Briggs, Frank H},
  journal={Physics reports},
  volume={433},
  number={4-6},
  pages={181--301},
  year={2006},
  publisher={Elsevier},
  doi = {https://doi.org/10.1016/j.physrep.2006.08.002},
  url = {https://doi.org/10.48550/arXiv.astro-ph/0608032},
  eprint = {arXiv:astro-ph/0608032},
}

@article{sigurdson2004dark,
  title={Dark-matter electric and magnetic dipole moments},
  author={Sigurdson, Kris and Doran, Michael and Kurylov, Andriy and Caldwell, Robert R and Kamionkowski, Marc},
  journal={Physical Review D—Particles, Fields, Gravitation, and Cosmology},
  volume={70},
  number={8},
  pages={083501},
  year={2004},
  publisher={APS},
  doi = {https://doi.org/10.1103/PhysRevD.73.089903},
  url = {https://doi.org/10.48550/arXiv.astro-ph/0406355},
  eprint = {arXiv:astro-ph/0406355},
}

@article{boehm2005constraints,
  title={Constraints on dark matter interactions from structure formation: Damping lengths},
  author={Boehm, Celine and Schaeffer, Richard},
  journal={Astronomy \& Astrophysics},
  volume={438},
  number={2},
  pages={419--442},
  year={2005},
  publisher={EDP Sciences},
  doi = {https://doi.org/10.1051/0004-6361%3A20042238},
  url = {https://doi.org/10.48550/arXiv.astro-ph/0410591},
  eprint = {arXiv:astro-ph/0410591},
}

@article{xu2021constraints,
  url = {https://arxiv.org/abs/2112.00707},
  eprint = {arXiv:2112.00707 },
  title={Constraints on GeV Dark Matter interaction with baryons, from a novel Dewar experiment},
  author={Xu, Xingchen and Farrar, Glennys R},
  year={2021}
}

@article{chen2002cosmic,
  title={Cosmic microwave background and large scale structure limits on the interaction between dark matter and baryons},
  author={Chen, Xuelei and Hannestad, Steen and Scherrer, Robert J},
  journal={Physical Review D},
  volume={65},
  number={12},
  pages={123515},
  year={2002},
  publisher={APS},
  doi = {https://doi.org/10.1103/PhysRevD.65.123515},
  url = {https://doi.org/10.48550/arXiv.astro-ph/0202496},
  eprint = {arXiv:astro-ph/0202496},
}

@article{short2022dark,
  url = {https://arxiv.org/abs/2203.16524},
  eprint = {arXiv:2203.16524},
  title={Dark matter-baryon scattering effects on temperature perturbations and implications for cosmic dawn},
  author={Short, Kathleen and Bernal, Jos{\'e} Luis and Boddy, Kimberly K and Gluscevic, Vera and Verde, Licia},
  year={2022}
}

@article{he2023s8,
  title={S8 tension in the context of dark matter--baryon scattering},
  author={He, Adam and Ivanov, Mikhail M and An, Rui and Gluscevic, Vera},
  journal={The Astrophysical Journal Letters},
  volume={954},
  number={1},
  pages={L8},
  year={2023},
  publisher={IOP Publishing},
  doi = {https://doi.org/10.3847/2041-8213/acdb63},
  url = {https://doi.org/10.48550/arXiv.2301.08260},
  eprint = {arXiv:2301.08260},
}

@article{mcdermott2011turning,
  title={Turning off the lights: How dark is dark matter?},
  author={McDermott, Samuel D and Yu, Hai-Bo and Zurek, Kathryn M},
  journal={Physical Review D—Particles, Fields, Gravitation, and Cosmology},
  volume={83},
  number={6},
  pages={063509},
  year={2011},
  publisher={APS},
  doi = {https://doi.org/10.1103/PhysRevD.83.063509},
  url = {https://doi.org/10.48550/arXiv.1011.2907},
  eprint = {arXiv:1011.2907},
}

@article{munoz2018insights,
  title={Insights on dark matter from hydrogen during cosmic dawn},
  author={Munoz, Julian B and Loeb, Abraham},
  journal={Nature},
  volume={557},
  publisher={Nature Publishing Group UK London},
  year={2018},
  doi = {https://doi.org/10.1038/s41586-018-0151-x},
  url = {https://doi.org/10.48550/arXiv.1802.10094},
  eprint = {arXiv:1802.10094},
}

@article{barkana2018signs,
  title = {Strong constraints on light dark matter interpretation of the EDGES signal},
  author = {Barkana, Rennan and Outmezguine, Nadav Joseph and Redigolo, Diego and Volansky, Tomer},
  journal = {Phys. Rev. D},
  volume = {98},
  issue = {10},
  pages = {103005},
  numpages = {11},
  year = {2018},
  month = {Nov},
  publisher = {American Physical Society},
  doi = {10.1103/PhysRevD.98.103005},
  url = {https://link.aps.org/doi/10.1103/PhysRevD.98.103005}
}

@article{liu2018implications,
  title={Implications of a 21-cm signal for dark matter annihilation and decay},
  author={Liu, Hongwan and Slatyer, Tracy R},
  journal={Physical Review D},
  volume={98},
  number={2},
  pages={023501},
  year={2018},
  publisher={APS},
  doi = {https://doi.org/10.1103/PhysRevD.98.023501},
  url = {https://doi.org/10.48550/arXiv.1803.09739},
  eprint = {arXiv:1803.09739},
}

@article{munoz201821,
  title={21-cm fluctuations from charged dark matter},
  author={Mu{\~n}oz, Julian B and Dvorkin, Cora and Loeb, Abraham},
  journal={Physical review letters},
  volume={121},
  number={12},
  pages={121301},
  year={2018},
  publisher={APS},
  doi = {https://doi.org/10.1103/PhysRevLett.121.121301},
  url = {https://doi.org/10.48550/arXiv.1804.01092},
  eprint = {arXiv:1804.01092},
}

@article{liu2019reviving,
  title={Reviving millicharged dark matter for 21-cm cosmology},
  author={Liu, Hongwan and Outmezguine, Nadav Joseph and Redigolo, Diego and Volansky, Tomer},
  journal={Physical Review D},
  volume={100},
  number={12},
  pages={123011},
  year={2019},
  publisher={APS},
  doi = {https://doi.org/10.1103/PhysRevD.100.123011},
  url = {https://doi.org/10.48550/arXiv.1908.06986},
  eprint = {arXiv:1908.06986},
}

@article{barkana2023anticipating,
  title={Anticipating a new physics signal in upcoming 21-cm power spectrum observations},
  author={Barkana, Rennan and Fialkov, Anastasia and Liu, Hongwan and Outmezguine, Nadav Joseph},
  journal={Physical Review D},
  volume={108},
  number={6},
  pages={063503},
  year={2023},
  publisher={APS},
  doi = {https://doi.org/10.1103/PhysRevD.108.063503},
  url = {https://doi.org/10.48550/arXiv.2212.08082},
  eprint = {arXiv:2212.08082},
}

@article{mosbech2023probing,
  title={Probing dark matter interactions with 21cm observations},
  author={Mosbech, Markus R and Boehm, Celine and Wong, Yvonne YY},
  journal={Journal of Cosmology and Astroparticle Physics},
  volume={2023},
  number={03},
  pages={047},
  year={2023},
  publisher={IOP Publishing},
  doi = {https://doi.org/10.1088/1475-7516/2023/03/047},
  url = {https://doi.org/10.48550/arXiv.2207.03107},
  eprint = {arXiv:2207.03107},
}

@article{benson2012galacticus,
  title={Galacticus: A semi-analytic model of galaxy formation},
  author={Benson, Andrew J},
  journal={New Astronomy},
  volume={17},
  number={2},
  pages={175--197},
  year={2012},
  publisher={Elsevier},
  doi = {https://doi.org/10.1016/j.newast.2011.07.004},
  url = {https://doi.org/10.48550/arXiv.1008.1786},
  eprint = {arXiv:1008.1786},
}

@article{li2018disentangling,
  title={Disentangling dark physics with cosmic microwave background experiments},
  author={Li, Zack and Gluscevic, Vera and Boddy, Kimberly K and Madhavacheril, Mathew S},
  journal={Physical Review D},
  volume={98},
  number={12},
  pages={123524},
  year={2018},
  publisher={APS},
  doi = {https://doi.org/10.1103/PhysRevD.98.123524},
  url = {https://doi.org/10.48550/arXiv.1806.10165},
  eprint = {arXiv:1806.10165},
}

@article{Rahimieh:2025lbf,
    author = "Rahimieh, Aryan and Parashari, Priyank and Gluscevic, Vera",
    title = "{Forecasting 21-cm power spectrum sensitivity to dark Matter-baryon scattering}",
    eprint = "2508.20507",
    archivePrefix = "arXiv",
    primaryClass = "astro-ph.CO",
    doi = "10.1093/mnras/staf1326",
    journal = "Mon. Not. Roy. Astron. Soc.",
    volume = "542",
    pages = "1605--1615",
    year = "2025"
}

@article{he2025bounds,
  title={Bounds on velocity-dependent dark matter-baryon scattering from large-scale structure},
  author={He, Adam and Ivanov, Mikhail M and An, Rui and Driskell, Trey and Gluscevic, Vera},
  journal={Journal of Cosmology and Astroparticle Physics},
  volume={2025},
  number={05},
  pages={087},
  year={2025},
  publisher={IOP Publishing},
  doi = {https://doi.org/10.1088/1475-7516/2025/05/087},
  url = {https://doi.org/10.48550/arXiv.2502.02636},
  eprint = {arXiv:2502.02636},
}

@article{buen2022cosmological,
  title={Cosmological constraints on dark matter interactions with ordinary matter},
  author={Buen-Abad, Manuel A and Essig, Rouven and McKeen, David and Zhong, Yi-Ming},
  journal={Physics Reports},
  volume={961},
  pages={1--35},
  year={2022},
  publisher={Elsevier},
  doi = {https://doi.org/10.1016/j.physrep.2022.02.006},
  url = {https://doi.org/10.48550/arXiv.2107.12377},
  eprint = {arXiv:2107.12377},
}

@article{creque2019direct,
  title={Direct millicharged dark matter cannot explain the EDGES signal},
  author={Creque-Sarbinowski, Cyril and Ji, Lingyuan and Kovetz, Ely D and Kamionkowski, Marc},
  journal={Physical Review D},
  volume={100},
  number={2},
  pages={023528},
  year={2019},
  publisher={APS},
  doi = {https://doi.org/10.1103/PhysRevD.100.023528},
  url = {https://doi.org/10.48550/arXiv.1903.09154},
  eprint = {arXiv:1903.09154},
}

@article{aprile2016low,
  title={Low-mass dark matter search using ionization signals in XENON100},
  author={Aprile, E and Aalbers, J and Agostini, F and Alfonsi, M and Amaro, FD and Anthony, M and Arneodo, F and Barrow, P and Baudis, L and Bauermeister, Boris and others},
  journal={Physical Review D},
  volume={94},
  number={9},
  pages={092001},
  year={2016},
  publisher={APS},
  doi = {https://doi.org/10.1103/PhysRevD.94.092001},
  url = {https://doi.org/10.48550/arXiv.1605.06262},
  eprint = {arXiv:1605.06262},
}

@article{agnes2018constraints,
  title={Constraints on sub-GeV dark-matter--electron scattering from the DarkSide-50 experiment},
  author={Agnes, Paolo and Albuquerque, Ivone Freire da Mota and Alexander, T and Alton, AK and Araujo, GR and Asner, DM and Ave, M and Back, HO and Baldin, B and Batignani, G and others},
  journal={Physical review letters},
  volume={121},
  number={11},
  pages={111303},
  year={2018},
  publisher={APS},
  doi = {https://doi.org/10.1103/PhysRevLett.121.111303},
  url = {https://doi.org/10.48550/arXiv.1802.06998},
  eprint = {arXiv:1802.06998},
}

@article{agnese2018first,
  title={First dark matter constraints from a SuperCDMS single-charge sensitive detector},
  author={Agnese, R and Aralis, T and Aramaki, T and Arnquist, IJ and Azadbakht, E and Baker, W and Banik, S and Barker, D and Bauer, DA and Binder, T and others},
  journal={Physical review letters},
  volume={121},
  number={5},
  pages={051301},
  year={2018},
  publisher={APS},
  doi = {https://doi.org/10.1103/PhysRevLett.121.051301},
  url = {https://doi.org/10.48550/arXiv.1804.10697},
  eprint = {arXiv:1804.10697},
}

@article{barak2020sensei,
  title={SENSEI: Direct-detection results on sub-GeV dark matter from a new skipper CCD},
  author={Barak, Liron and Bloch, Itay M and Cababie, Mariano and Cancelo, Gustavo and Chaplinsky, Luke and Chierchie, Fernando and Crisler, Michael and Drlica-Wagner, Alex and Essig, Rouven and Estrada, Juan and others},
  journal={Physical Review Letters},
  volume={125},
  number={17},
  pages={171802},
  year={2020},
  publisher={APS},
  doi = {https://doi.org/10.1103/PhysRevLett.125.171802},
  url = {https://doi.org/10.48550/arXiv.2004.11378},
  eprint = {arXiv:2004.11378},
}

@article{bringmann2019novel,
  title={Novel direct detection constraints on light dark matter},
  author={Bringmann, Torsten and Pospelov, Maxim},
  journal={Physical review letters},
  volume={122},
  number={17},
  pages={171801},
  year={2019},
  publisher={APS},
  doi = {https://doi.org/10.1103/PhysRevLett.122.171801},
  url = {https://doi.org/10.48550/arXiv.1810.10543},
  eprint = {arXiv:1810.10543},
}

@article{cappiello2019strong,
  title={Strong new limits on light dark matter from neutrino experiments},
  author={Cappiello, Christopher V and Beacom, John F},
  journal={Physical Review D},
  volume={100},
  number={10},
  pages={103011},
  year={2019},
  publisher={APS},
  doi = {https://doi.org/10.1103/PhysRevD.100.103011},
  url = {https://doi.org/10.48550/arXiv.1906.11283},
  eprint = {arXiv:1906.11283},
}

\end{document}